\documentclass[aps,prb,10pt,superscriptaddress,reprint]{revtex4-1}

\usepackage{amsmath}
\usepackage{graphicx}
\usepackage{multirow}
\usepackage{dcolumn}
\newcolumntype{.}{D{.}{.}{1}}
\usepackage{bm}
\usepackage[usenames]{color}
\usepackage{hyperref}
\hypersetup{
 pdfnewwindow=true, colorlinks=true,
 linkcolor=blue, anchorcolor=blue,
 citecolor=blue, filecolor=blue,
 menucolor=blue, urlcolor=blue}
\usepackage{breakurl}

\begin{document}

\title{Insulating titanium oxynitride for visible light photocatalysis}

\author{Yuta Aoki} 
\altaffiliation{Current address: Center for Materials Research by Information Integration, National Institute for Materials Science, 1-2-1 Sengen, Tsukuba, Ibaraki 305-0047, Japan}
\email{AOKI.Yuta@nims.go.jp}
\affiliation{Department of Physics, Tokyo Institute of Technology,
  2-12-1 Oh-okayama, Meguro-ku, Tokyo 152-8551, Japan}
\affiliation{International Education and Research Center of Science,
  Tokyo Institute of Technology, 2-12-1 Oh-okayama, Meguro-ku, Tokyo
  152-8551, Japan}
\affiliation{Department of Physics, University of California,
  Berkeley, California 94720, USA}

\author{Sinisa Coh}
\affiliation{Department of Physics, University of California,
  Berkeley, California 94720, USA}
\affiliation{Materials Sciences Division, Lawrence Berkeley National
  Laboratory, Berkeley, California 94720, USA}
\affiliation{Department of Mechanical Engineering, Materials Science and Engineering, University of California Riverside, Riverside, CA 92521, USA}

\author{Steven G. Louie}
\affiliation{Department of Physics, University of California,
  Berkeley, California 94720, USA}
\affiliation{Materials Sciences Division, Lawrence Berkeley National
  Laboratory, Berkeley, California 94720, USA}

\author{Marvin L. Cohen}
\affiliation{Department of Physics, University of California,
  Berkeley, California 94720, USA}
\affiliation{Materials Sciences Division, Lawrence Berkeley National
  Laboratory, Berkeley, California 94720, USA}

\author{Susumu Saito}
\affiliation{Department of Physics, Tokyo Institute of Technology,
  2-12-1 Oh-okayama, Meguro-ku, Tokyo 152-8551, Japan}
\affiliation{International Research Center for Nanoscience and Quantum
  Physics, Tokyo Institute of Technology, 2-12-1 Oh-okayama,
  Meguro-ku, Tokyo 152-8551, Japan}
\affiliation{Materials Research Center for Element Strategy, Tokyo
  Institute of Technology, 4259 Nagatsutacho, Midori-ku, Yokohama,
  Kanagawa 226-8503, Japan}

\date{\today}


\begin{abstract}
We propose a systematic approach to obtain various forms of insulating titanium oxynitrides Ti$_{n}$N$_{2}$O$_{2n-3}$ and 
we conduct a detailed study on its $n=2$ case, Ti$_2$N$_2$O.
We study the energetics and the electronic structures of Ti$_2$N$_2$O and compare these results
with those of pristine and nitrogen-doped TiO$_2$ within the framework of the density-functional theory (DFT)
and the GW approximation. 
We find that Ti$_2$N$_2$O is semiconducting  with the calculated band-gap of 1.81~eV,
which is
significantly smaller than those of pristine TiO$_2$ rutile (3.14~eV) or anatase (3.55~eV). 
Furthermore, the reduction of the band-gap  of Ti$_2$N$_2$O is realized
not by lowering of the conduction-band minimum but by rising the valence-band maximum.  Therefore
the proposed Ti$_2$N$_2$O has suitable band-edge alignment for water-splitting photocatalysis. 
Finally, total energy calculations indicate that Ti$_2$N$_2$O is potentially easier to synthesize than nitrogen-doped TiO$_2$.
Based on these results, we propose Ti$_2$N$_2$O as a promising visible-light photocatalytic material. 
\end{abstract}

\maketitle

\section{Introduction}

Titanium dioxide TiO$_2$ is extensively studied as a functional material
with the potential of being used in various technological applications
such as photocatalysis, photovoltaics, and oxide electronics.\cite{rmp,review1,review2,review3}
One of the most interesting applications is the photocatalysis of water into hydrogen and oxygen gas.\cite{fujishima,review4,review5,renew} 
However, the experimental optical band gap of TiO$_2$ is 3.0~eV in rutile and 3.2~eV in anatase phase. \cite{PCM,Minoura,Tang,Tang2,KG,Hosoka}
Therefore pristine TiO$_2$ is not photoactive under a large part of the solar spectrum.\cite{review4,renew}
However, to improve water splitting efficiency, it is not enough to just reduce the band gap.  
TiO$_2$ must maintain appropriate band energy alignment with water oxidation and reduction levels.
While the conduction band in TiO$_2$ is already aligned well with water reduction level, 
its valence band is too low in energy. 
In order to optimize the photocatalytic performance, the band-gap reduction in titanium oxide must come from an
upward energy shift of the valence band, not from a downward shift of the conduction band. \cite{renew, aoki1}

Several methods such as impurity doping and forming solid solutions have been proposed in 
earlier reports to improve photocatalytic activity of TiO$_2$ under visible light.\cite{renew,review6,reivew7,review8,review9}
As for the solid-solution method, the TiO$_2$-ZrO$_2$ solid solution was reported to have the absorption tail in the visible-light region.\cite{solution2} Also, the reduction of the band gap was theoretically predicted for 
the TaON-TiO$_2$ solution.\cite{solution4}
On the other hand, nitrogen doping was shown to add the absorption tail in the visible-light region by Asahi et al.\cite{asahi1} 
and this was followed with many other experimental and theoretical
studies.\cite{asahi2,Irie,Lindgren,Torres,Mrowetz,Kobayakawa,Spada,Wu,Yang,Li1,Li2,hitosugi,nakano,matsui,anpo,Gole1,Gole2,aoki1,aoki2}
The effectiveness of the nitrogen doping is derived
from the fact that atomic $2p$ nitrogen states are higher in energy than atomic $2p$
oxygen states.  
Therefore nitrogen doping raises the valence band of
TiO$_2$ as it is formed primarily from oxygen $2p$ states.\cite{Taga,Mo,Nakai}

However, nitrogen doping into TiO$_2$ does not give a new peak in the visible-light region. Instead, it
just adds a 
shoulder to the original peak in the ultraviolet region.\cite{asahi2,Irie,Kobayakawa,Wu,Yang} 
Most likely, 
this happens because
the nitrogen concentration is so small that the absorption via the nitrogen-induced states only occurs around the nitrogen dopants. 
Therefore, to enhance visible-light absorption with nitrogen doping, it
is necessary to increase the nitrogen concentration level. 
To the best of our knowledge, the highest reported 
nitrogen concentration in TiO$_2$ is 15~\%.\cite{hitosugi} 
Several studies suggested that oxygen vacancies play a
crucial role in stabilizing nitrogen substitution,\cite{Ov1,Ov2} as is expected from
the electron-counting rule.
Since concentration of oxygen vacancies in native TiO$_2$ such as
rutile or anatase is relatively low, there is a natural limit to
the amount of nitrogen one can substitute into TiO$_2$.

Here we propose an alternative way to achieve much higher nitrogen 
concentrations in titanium oxide systems.
Instead of focusing on TiO$_2$ stoichiometry with a limited number of
native oxygen vacancies, we consider here ordered compounds with
composition Ti$_n$O$_{2n-1}$, where $n$ is any positive integer.  These
systems can be viewed as ordered oxygen-deficient variants of TiO$_2$
with a large concentration of oxygen vacancies.  Since oxygen
vacancies stabilize nitrogen substitutions, one can expect that these
systems could achieve much higher nitrogen concentrations than
TiO$_2$.  The compound with $n=2$ has the well known
corundum structure as its ground state\cite{ti2o3-c1,ti2o3-c2,ti2o3-c3} while the structures of $n \ge 3$ compounds are known as Magn\'{e}li
phases. \cite{magneli1,magneli2,magneli3} 

Since the nominal titanium valency in Ti$_n$O$_{2n-1}$ is less than +IV,
one might expect them to be metallic due to partially occupied
$d$-levels on titanium.  However, it is found experimentally that
corundum Ti$_2$O$_3$ and several Magn\'{e}li phases exhibit the
metal-insulator
transition\cite{ti2o3-1,ti2o3-2,ti2o3-3,ti2o3-4,ti2o3-9,MITreview,ti3o5-1,ti3o5-2,ti4o7-1,ti4o7-2,tio}
with a small charge-ordering gap one or two orders of magnitude less than in TiO$_2$.\cite{co1,co2,ti4o7-3} 
(The values of the charge-order gap in these systems are between 27 and 250~meV. \cite{ti2o3-2,ti2o3-3,ti2o3-5,ti2o3-6,ti2o3-7,
ti2o3-8,ti4o7-3,ti4o7-4,ti4o7-5}) 
Therefore, to
achieve large-gap insulating state without charge ordering, one needs
to replace two out of $2n-1$ oxygen atoms with nitrogen, thus achieving
composition
$$\text{Ti}_n\text{N}_2\text{O}_{2n-3} \text{\quad ($n\ge 2$)}$$ 
restoring the highly stable +IV valency of titanium. 
We propose this as a purely insulating titanium oxynitride. 
It is highly notable that this approach has high controllability over electronic properties since 
one can tune the nitrogen concentration while keeping the titanium valency +IV by choosing the appropriate value for $n$.

Nitrogen substitution into Ti$_n$O$_{2n-1}$ systems 
was first reported by Hyett and his co-workers for the $n=3$ case (Ti$_{3-\delta}$O$_4$N, $0.06<\delta<0.25$)\cite{oxynitride1,oxynitride3}
and later extended to larger $n$ ($5\le n\le 8$) cases by Mikami and Ozaki (Ti$_n$(O,N)$_{2n-1}$).\cite{oxynitride2}
However, the resistivity of their samples is quite low (10--1000 $\mu$$\Omega$m). 
Therefore these samples are likely quite far away from the ideal Ti$_n$N$_2$O$_{2n-3}$ configuration. 
As far as we know, insulating solid material with nominal Ti$_n$N$_2$O$_{2n-3}$ configuration has not been synthesized up to now
while there is an experimental report on the existence of molecules with the composition Ti$_2$N$_2$O, 
which corresponds to the $n=2$ case. \cite{Ti2N2Omol}
On the other hand, theoretical prediction of Ti$_n$N$_2$O$_{2n-3}$ solid compounds has been done for the $n=3$ case 
(Ti$_3$N$_2$O$_3$) by Wu and his co-workers via a high throughput screening method within the density functional theory (DFT).\cite{oxynitride4}
They predicted Ti$_3$N$_2$O$_3$ based on the structure of Ta$_3$N$_5$,\cite{Ta3N5} which is almost isomorphic to the $\alpha$-Ti$_3$O$_5$ structure. 
Within the $\Delta$-sol method they obtained the band gap of 2.37~eV with a favorable band-edge position for water splitting.\cite{deltasol}

In this paper, we focus on the oxynitride Ti$_2$N$_2$O ($n=2$ case) with
corundum-like structure since it has the highest nitrogen
concentration among all $n \ge 2$.  However, our conclusions might
extend to Magn\'{e}li phases ($n \ge 3$) as well.  In addition, we
study nitrogen-doped TiO$_2$ so that we can compare the nitrogen
substitutional energy and the electronic structure with those of
Ti$_2$N$_2$O.

\section{Methods}

Our electronic-structure calculations are based on the framework of the
density-functional theory (DFT).\cite{HK,KS} We also employ the GW
approach\cite{hedin,hybertsen-louie-1,hybertsen-louie-2} to compute
quasiparticle band structures of pristine TiO$_2$. 

In DFT calculations we use the generalized gradient approximation
(GGA)\cite{GGAperdew} with the Perdew-Burke-Ernzerhof (PBE)
exchange-correlation functional.\cite{PBE} We expand the Kohn-Sham
orbitals in plane-wave expansion with the cut-off energy of 200 Ry.
We employ the pseudopotential method with the Troullier-Martins
norm-conserving pseudopotentials. \cite{TM}  We generated
pseudopotentials for Ti, O, and N with the charged Ti$^{4+}$ and
neutral O and N configurations.
In generating the pseudopotential for Ti, the $3s$ and $3p$ semicore
states are treated as valence states and the cutoff radii for the
$3s$, $3p$, and $3d$ states are set to be 0.90~$a_{\rm B}$, 0.90~$a_{\rm B}$,
 and 1.00~$a_{\rm B}$, respectively, where $a_{\rm B}$ is
the Bohr radius.  In generating the pseudopotentials for O and N, the
cutoff radii both for the $2s$ and $2p$ states are set to be 1.05
$a_{\rm B}$.  The k-points sampling is performed using the
Monkhorst-Pack method\cite{monk} with 4$\times$4$\times$6,
4$\times$4$\times$2, and 6$\times$6$\times$2 grids for the primitive
unit cell of rutile TiO$_2$, and the conventional unit cells of anatase TiO$_2$ 
and corundum Ti$_2$O$_3$, respectively.  For metallic systems, we use the
Methfessel-Paxton first-order spreading with the searing of 0.01~Ry in
the Brillouin-zone integration. \cite{mp}  The structural optimization
is performed based on the Wentzcovitch damped molecular
dynamics. \cite{Wen}  The electronic-structure calculation within the
DFT is performed using the Quantum ESPRESSO package. \cite{QE}

Quasiparticle band structures are calculated by the
one-shot GW
approach\cite{hedin,hybertsen-louie-1,hybertsen-louie-2} within the
generalized plasmon-pole (GPP) model. \cite{gpp} As the starting point
for the one-shot GW calculation, we use the DFT results
obtained through the above-mentioned DFT methodology. The number of
the calculated bands within the DFT calculation is 1000 per TiO$_2$ formula unit.
The frequency cut-off for the dielectric matrix calculation is set to be 40~Ry. 
These parameters for the GW calculation were confirmed to give
well converged results in a recent GW study for rutile TiO$_2$. \cite{Andrei}
The GW calculation is
performed using the BerkeleyGW package. \cite{BerkeleyGW}

\section{Structure}

In this section, we discuss structural models of nitrogen-substituted TiO$_2$ and
Ti$_2$O$_3$ used in our study.

\subsection{Nitrogen substituted TiO$_2$}

\begin{figure}[htbp]
\begin{center}
\begin{tabular}{ccc}
(a) & (b) & \\
\includegraphics[scale=0.1,clip]{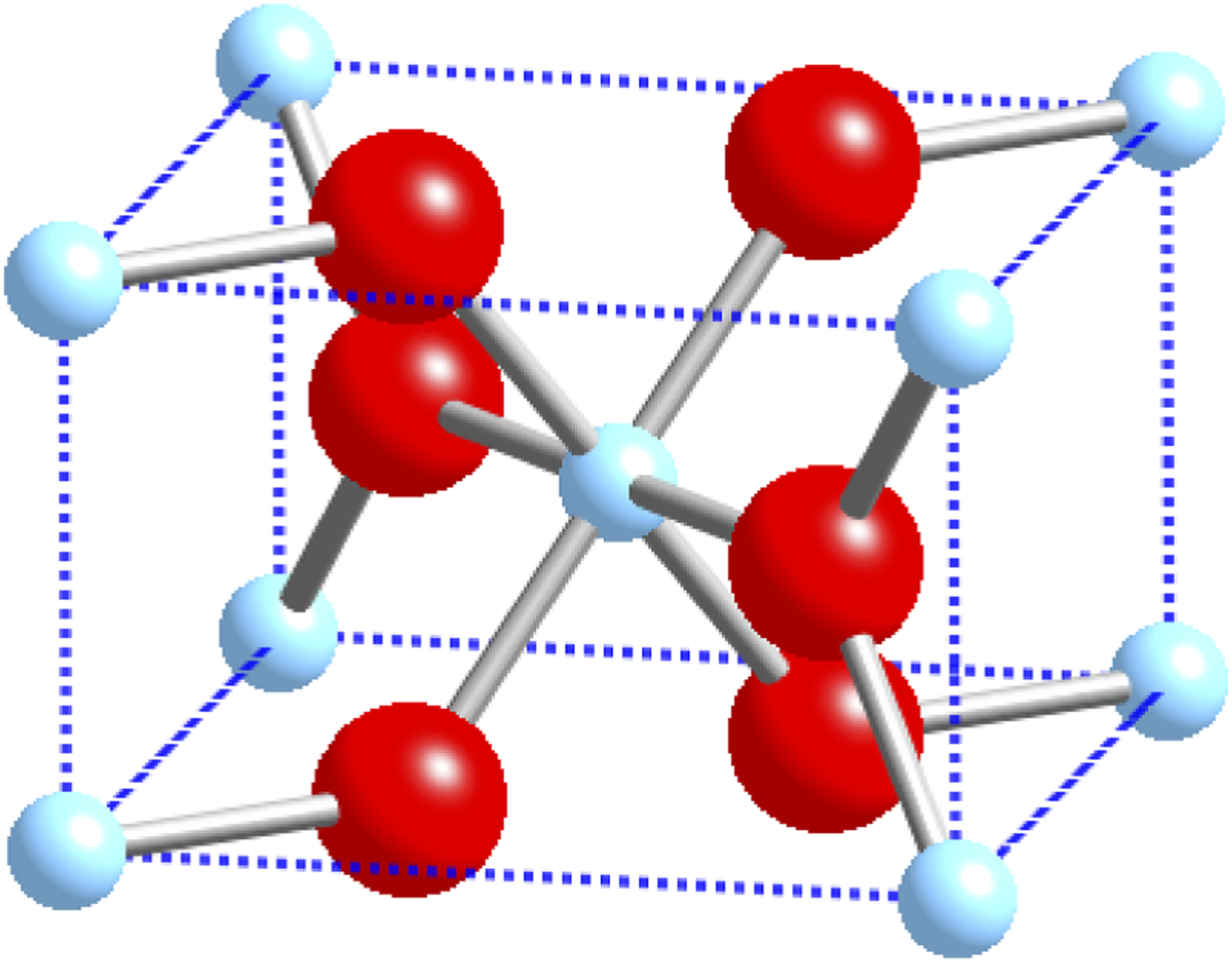} &
\includegraphics[scale=0.14,clip]{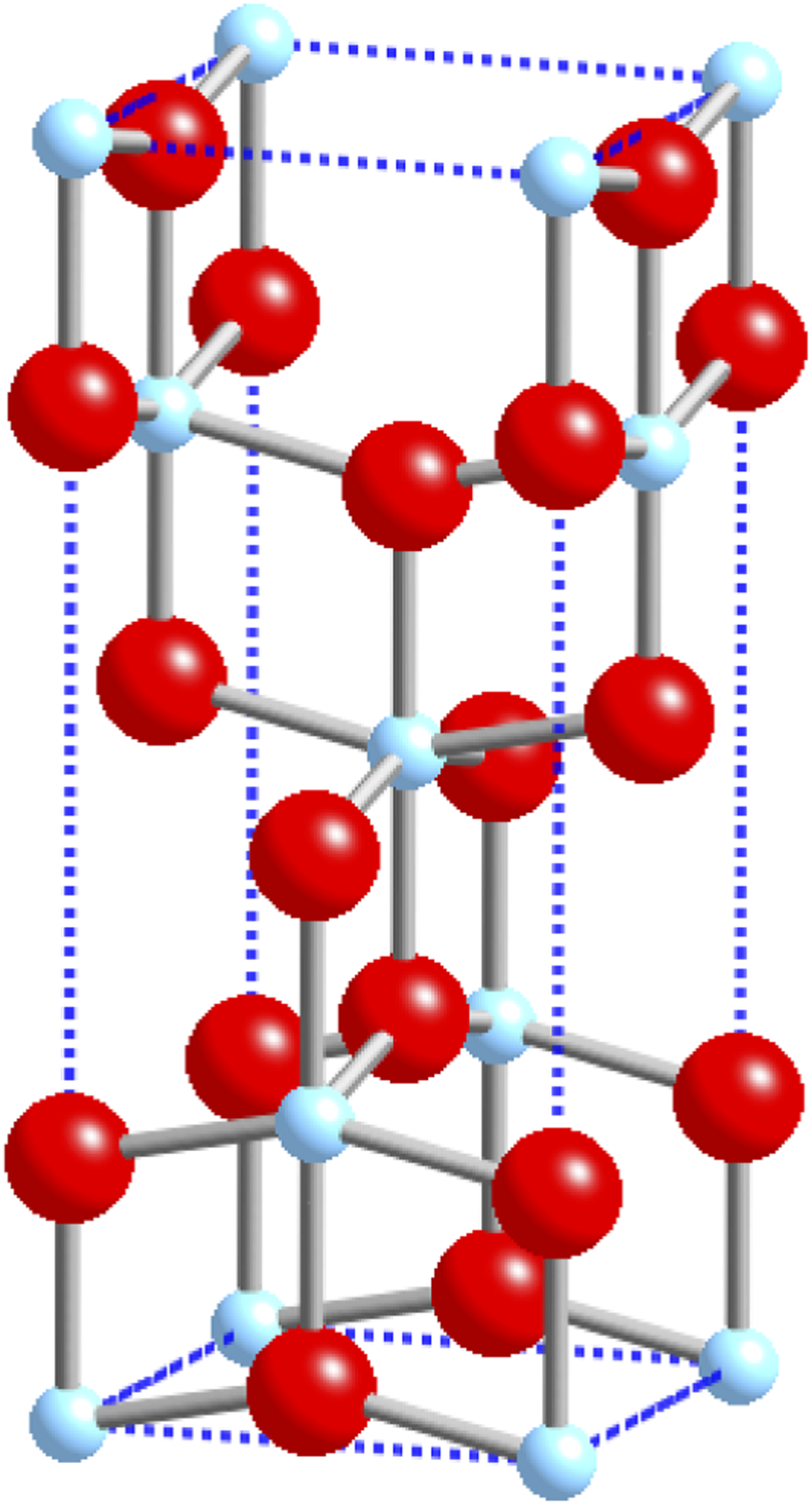} &
\includegraphics[scale=0.03]{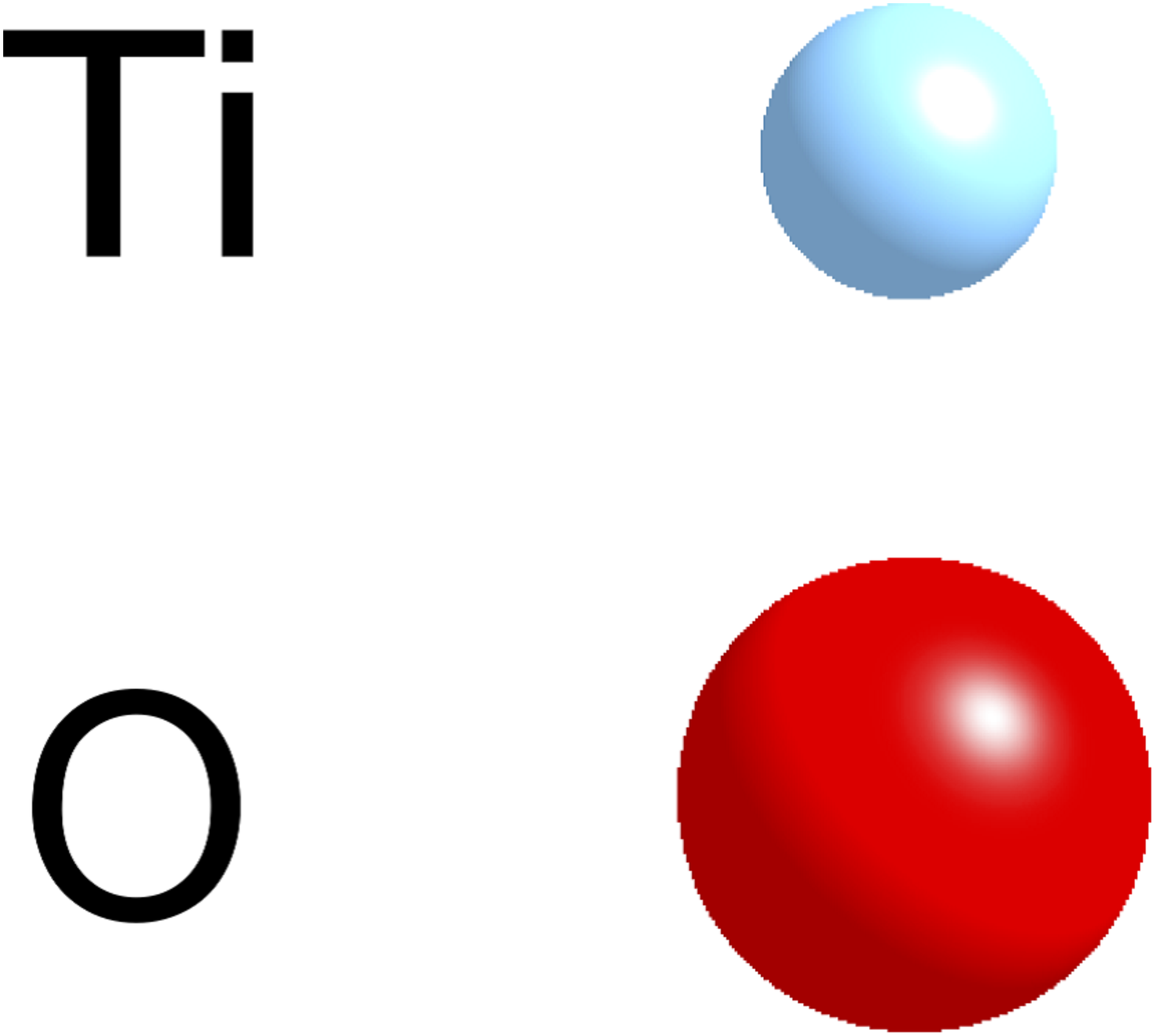}
\end{tabular}
\caption{(Color online) (a): The unit cell of rutile. (b): The conventional unit cell
  of anatase.}\label{pristinestructure}
\end{center}
\end{figure}

Figure~\ref{pristinestructure} shows the crystal structure of pristine
rutile and anatase TiO$_2$, which are representative phases among
TiO$_2$ polymorphs. Rutile has a primitive tetragonal
structure with the space group $P4_2/mnm\ (D_{4h}^{14})$.  Its unit
cell contains two TiO$_2$ formula units. Anatase has a body-centered
tetragonal structure with the space group $I4_1/amd\ (D^{19}_{4h})$.  Its
primitive unit cell also contains only two TiO$_2$ units.

\begin{figure}[htbp]
\begin{center}
\begin{tabular}{ccc}
(a) & (b) & \\
\includegraphics[scale=0.11]{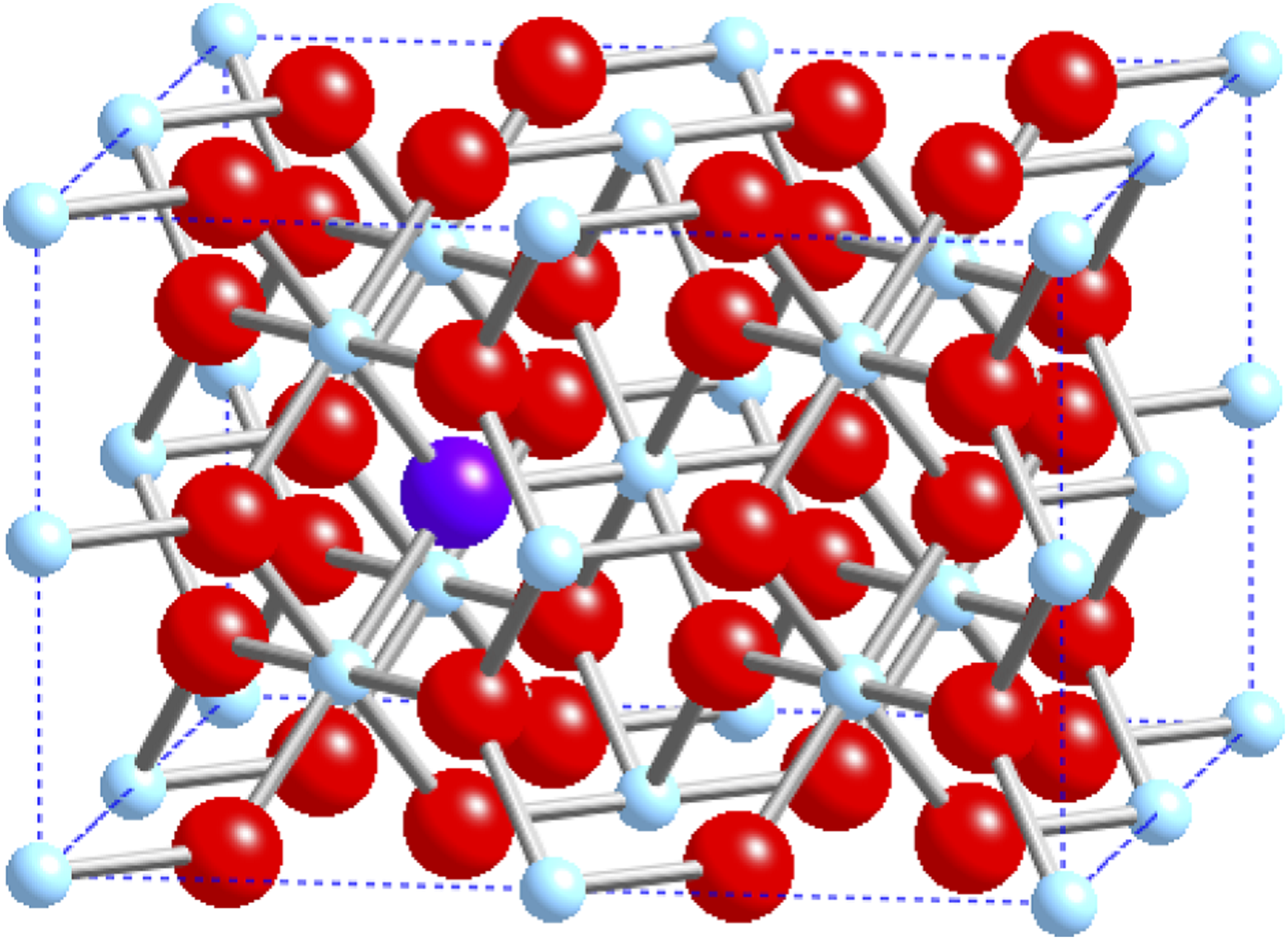}&
\includegraphics[scale=0.12]{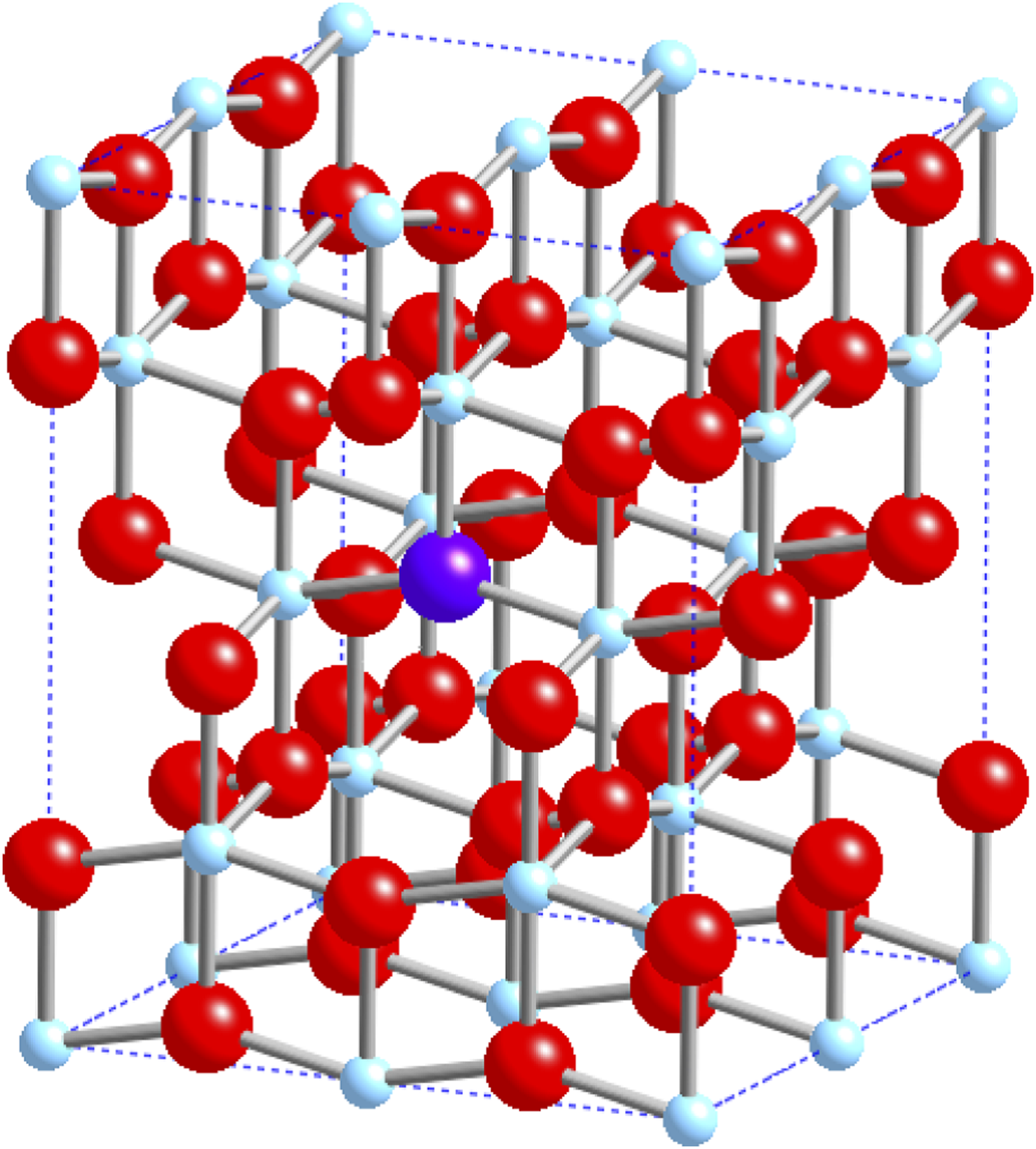}& 
\includegraphics[scale=0.03]{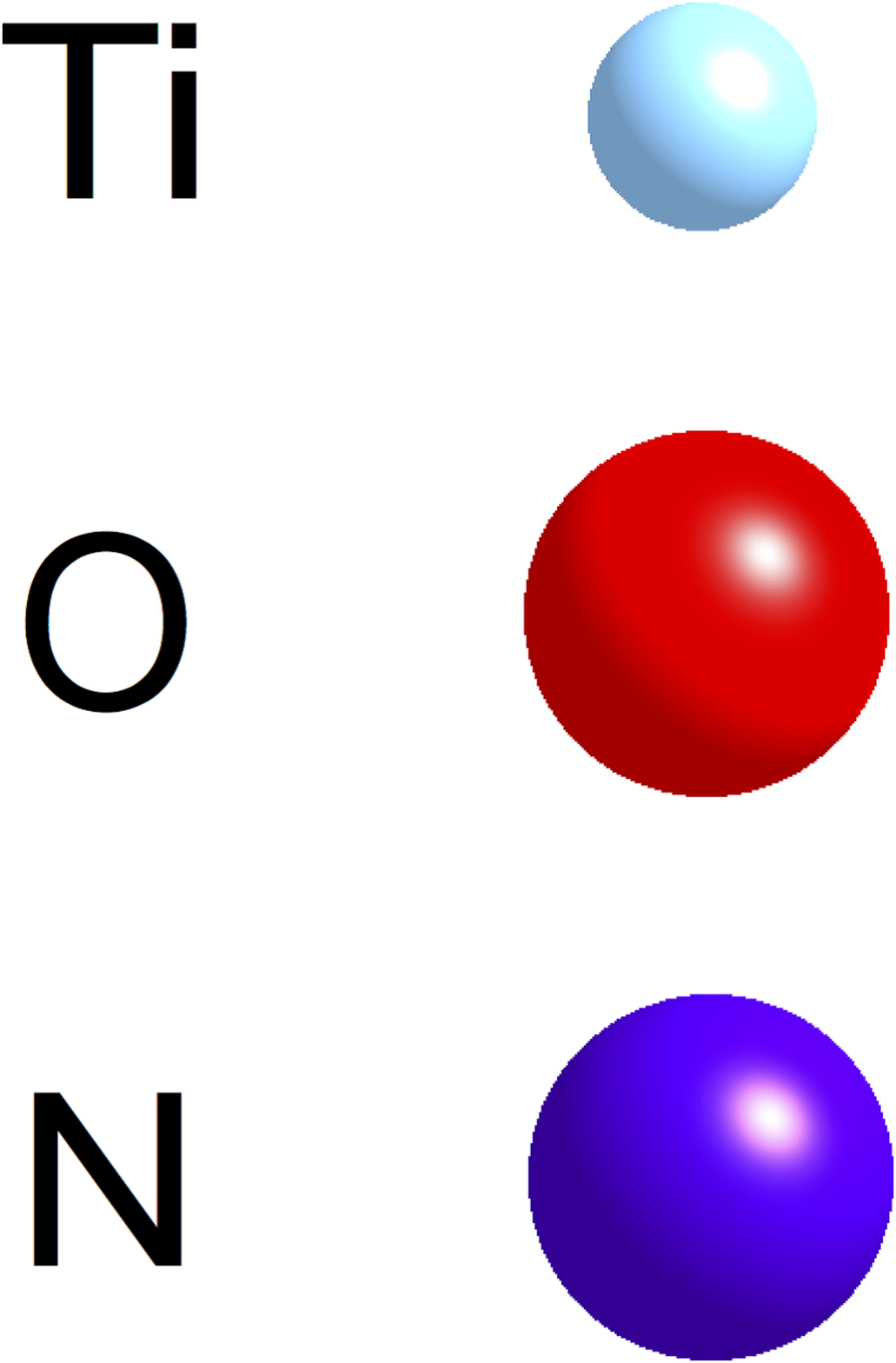}\\
\end{tabular}
\caption{(Color online) (a): Structural model for nitrogen-doped rutile. One oxygen
  atom is replaced by one nitrogen atom in the 2$\times$2$\times$2
  supercell of rutile. (b): Structural model for nitrogen-doped
  anatase. One oxygen atom is replaced by one nitrogen atom in the
  2$\times$2$\times$1 supercell of the conventional unit cell of
  anatase.}\label{NTi-st}
\end{center}
\end{figure}

All oxygen sites in rutile and anatase are crystallographically
equivalent.  Therefore, there is only one symmetry equivalent way to
substitute a single oxygen atom with a nitrogen atom in these
structures.  In our calculation we replace one out of thirty-two oxygen atoms
with nitrogen, both in rutile and anatase phase.  After this
substitution we fully relax the internal structural coordinates and
lattice constants.
In the case of rutile this requires a 2$\times$2$\times$2 supercell
and in the case of anatase 2$\times$2$\times$1 supercell of its
conventional unit cell.  These supercells are shown in
Fig.~\ref{NTi-st}.

\subsection{Nitrogen substituted Ti$_2$O$_3$: Ti$_2$N$_2$O}

\begin{figure}[htbp]
\begin{center}
\begin{tabular}{ccc}
(a) & (b) \\ 
\includegraphics[scale=0.105]{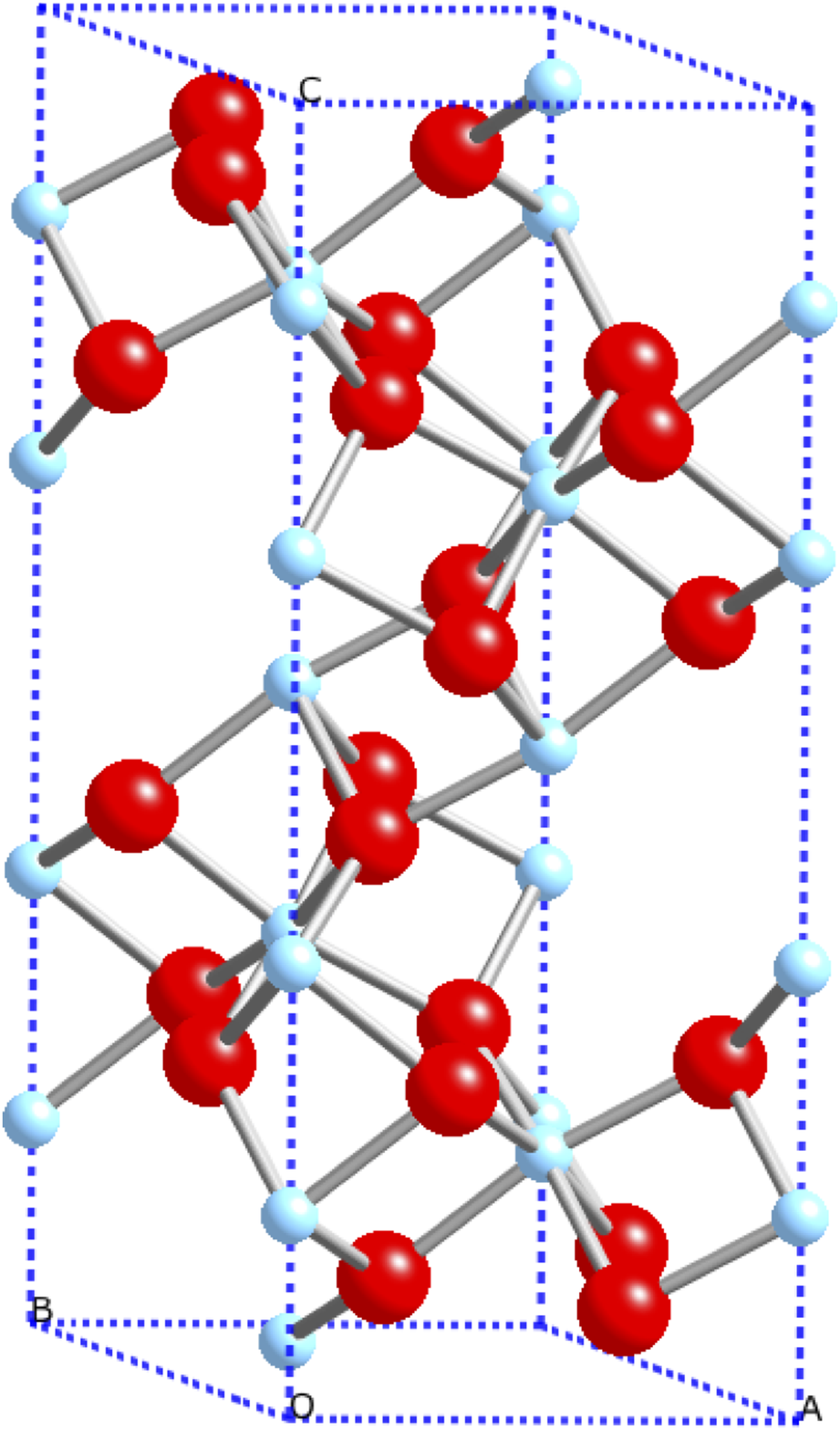} &
\includegraphics[scale=0.105]{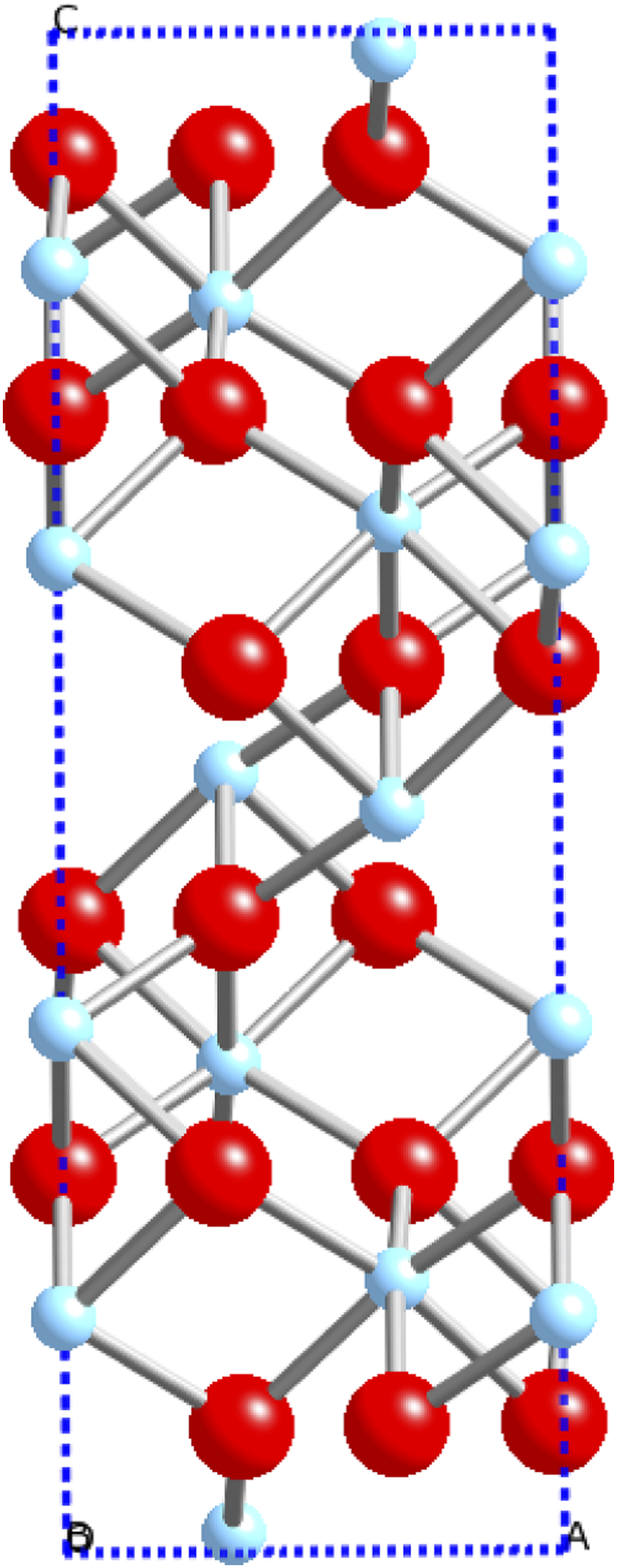} & \\
(c) & (d) \\
\includegraphics[scale=0.105]{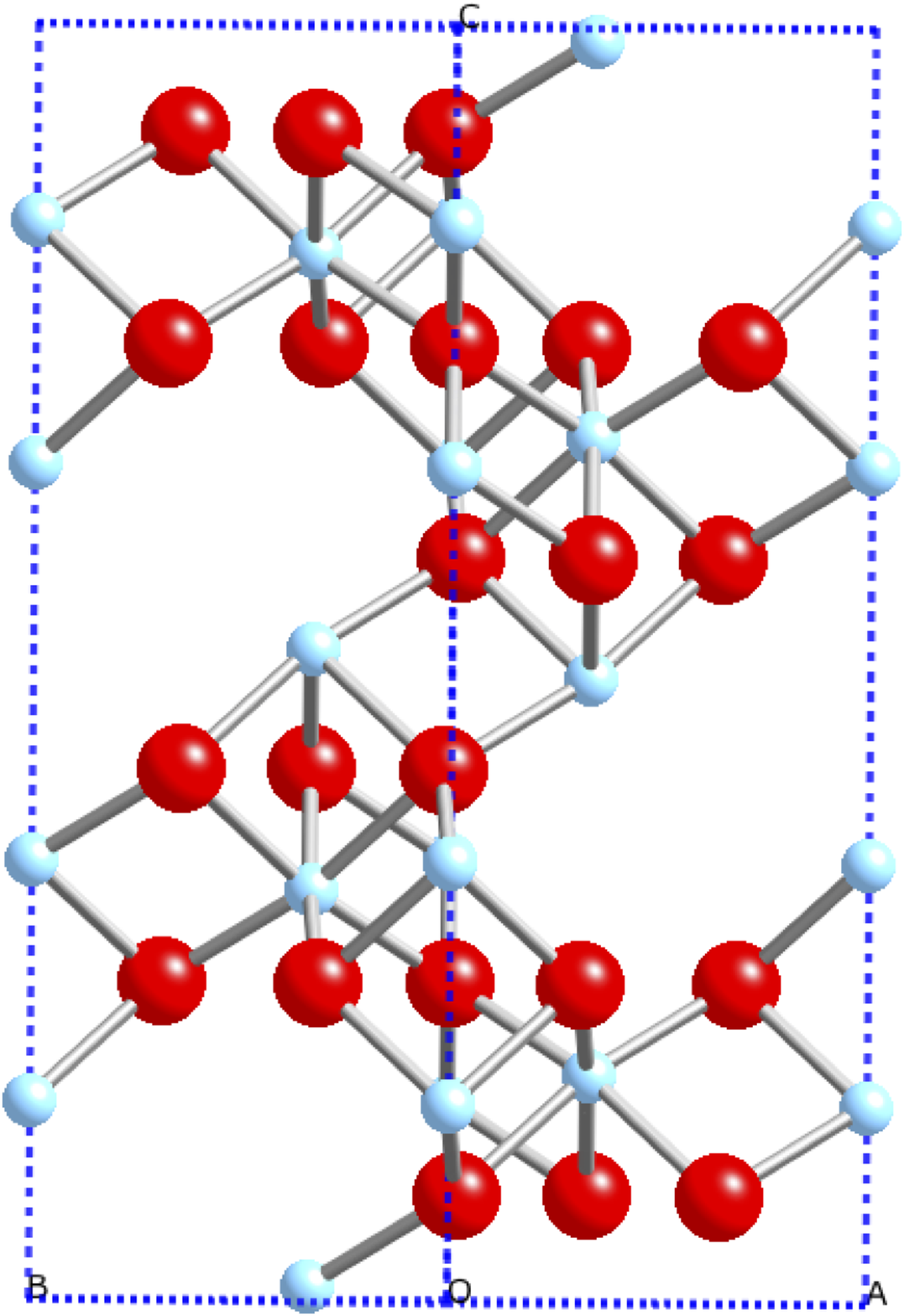} &
\includegraphics[scale=0.105]{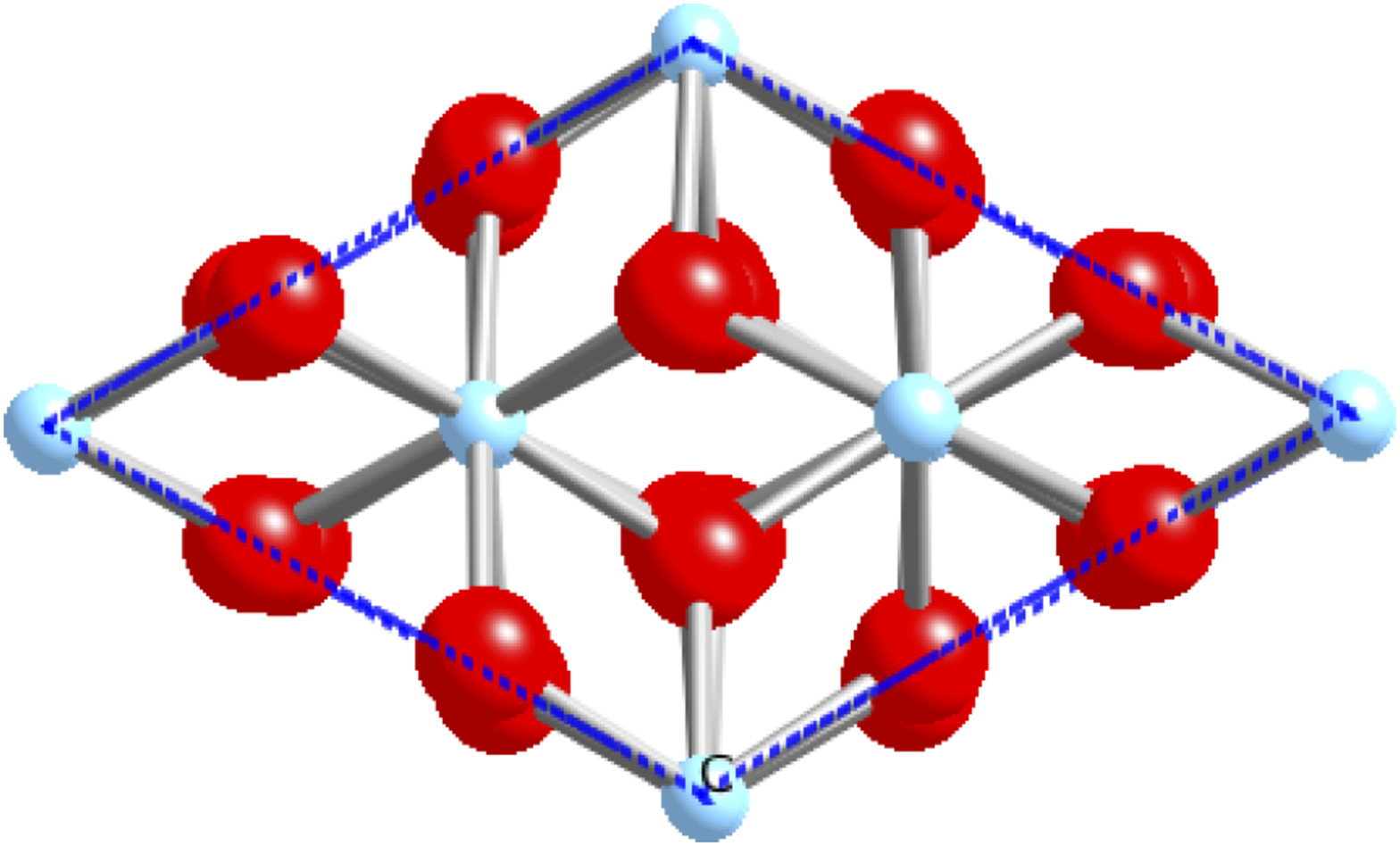} & 
\includegraphics[scale=0.03]{Ti-O_legend_1.eps}
\end{tabular}
\caption{(Color online) The conventional unit cell of corundum
  Ti$_2$O$_3$ are shown in four ways. (a): Bird-eye view. (b): Projection in the direction of ${\mathbf b}$. (c): Projection in the direction of $\mathbf{a} + \mathbf{b}$. (d): Projection in the direction of $-\mathbf{c}$ (top view).}\label{Ti2O3-st}
\end{center}
\end{figure}

The crystal structure of corundum Ti$_2$O$_3$ is shown in
Fig.~\ref{Ti2O3-st}. 
The corundum structure has rhombohedral symmetry with the space group $R\bar{3}c$ ($D^6_{3d}$) 
and its primitive unit cell contains two Ti$_2$O$_3$ units.

\begin{figure}[htbp]
\begin{center}
\begin{tabular}{ccc}
(a): Ti$_2$N$_2$O-I & (b): Ti$_2$N$_2$O-II &(c): Ti$_2$N$_2$O-III \\
\includegraphics[width=1in]{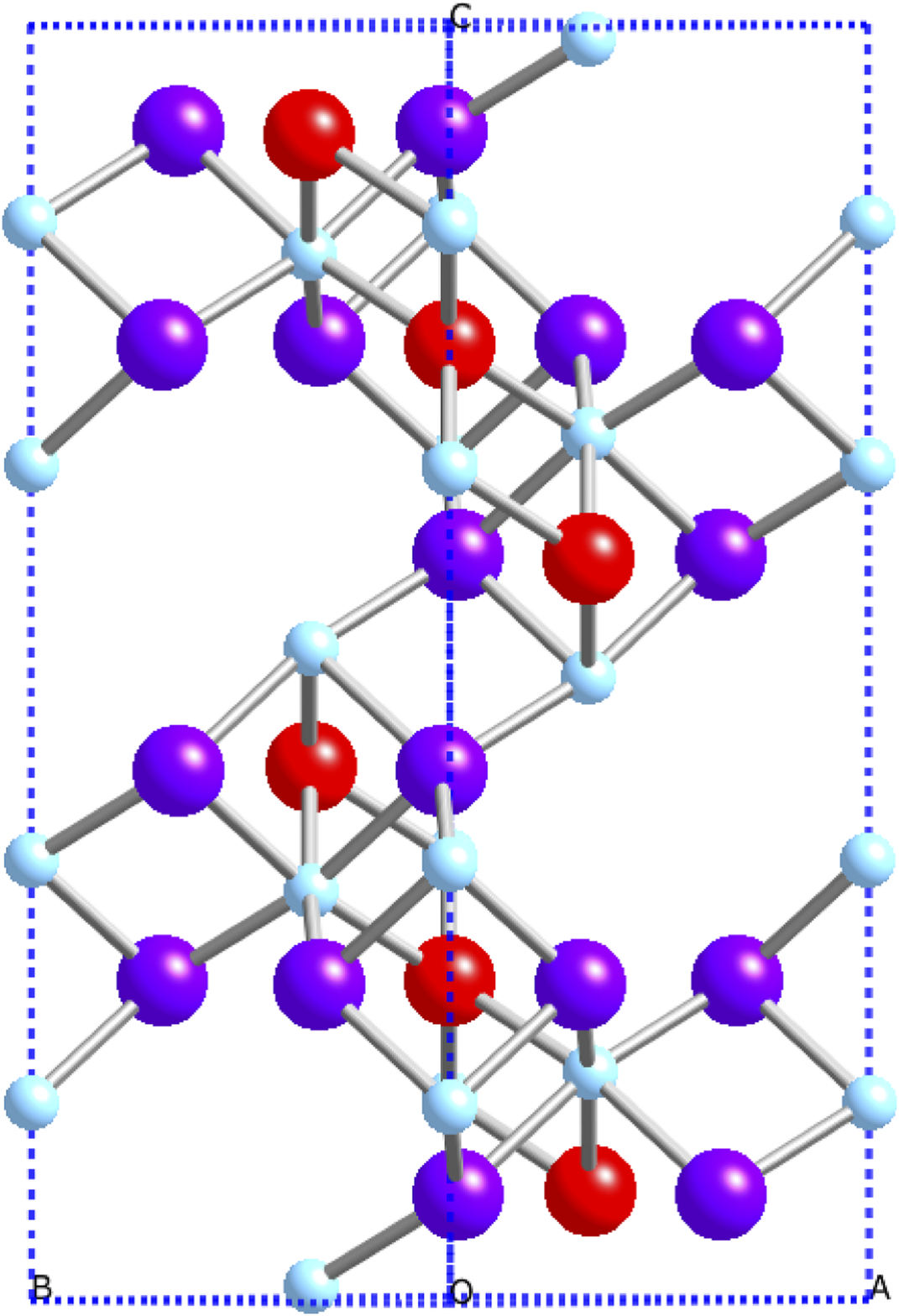} &
\includegraphics[width=1in]{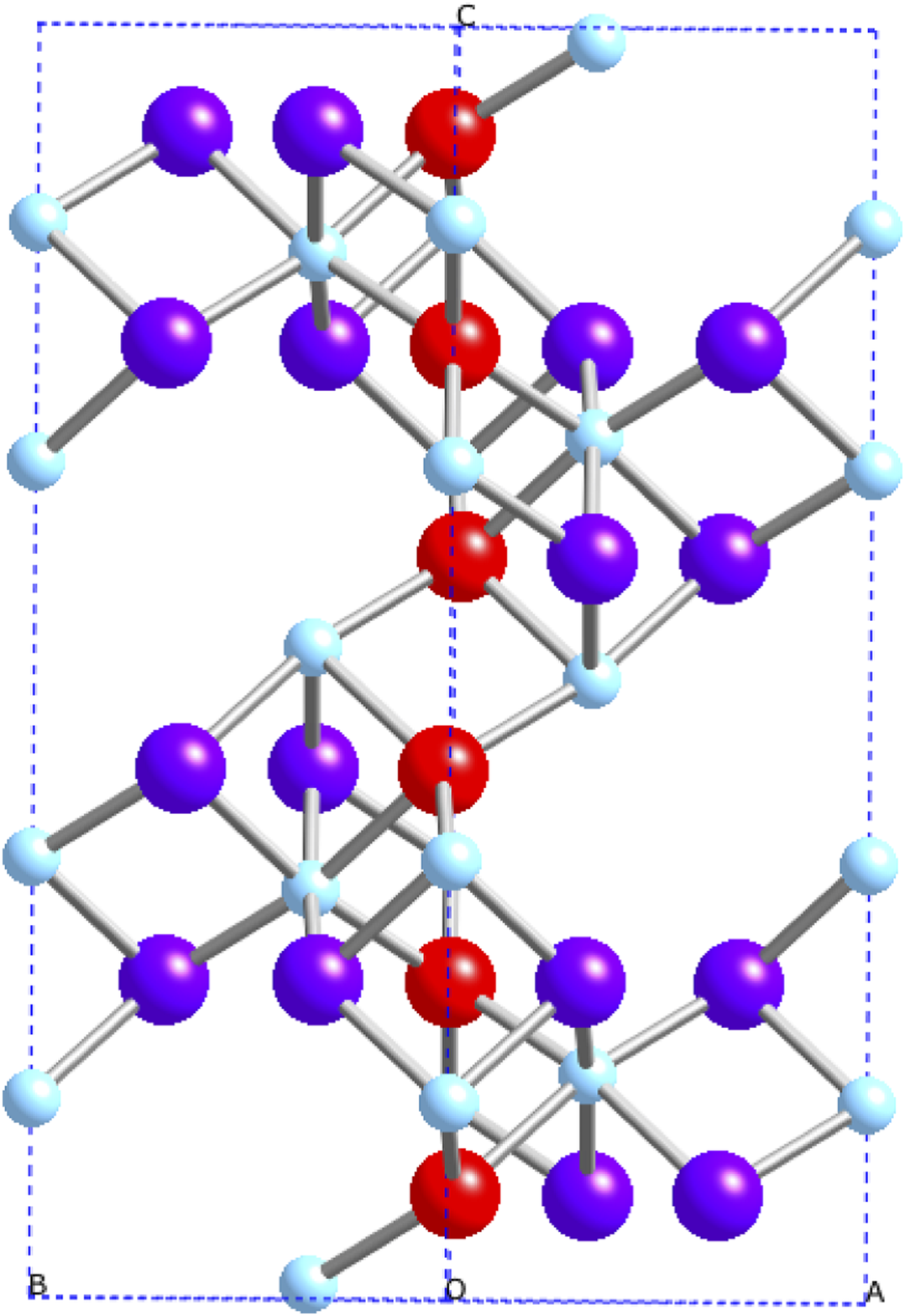} &
\includegraphics[width=1in]{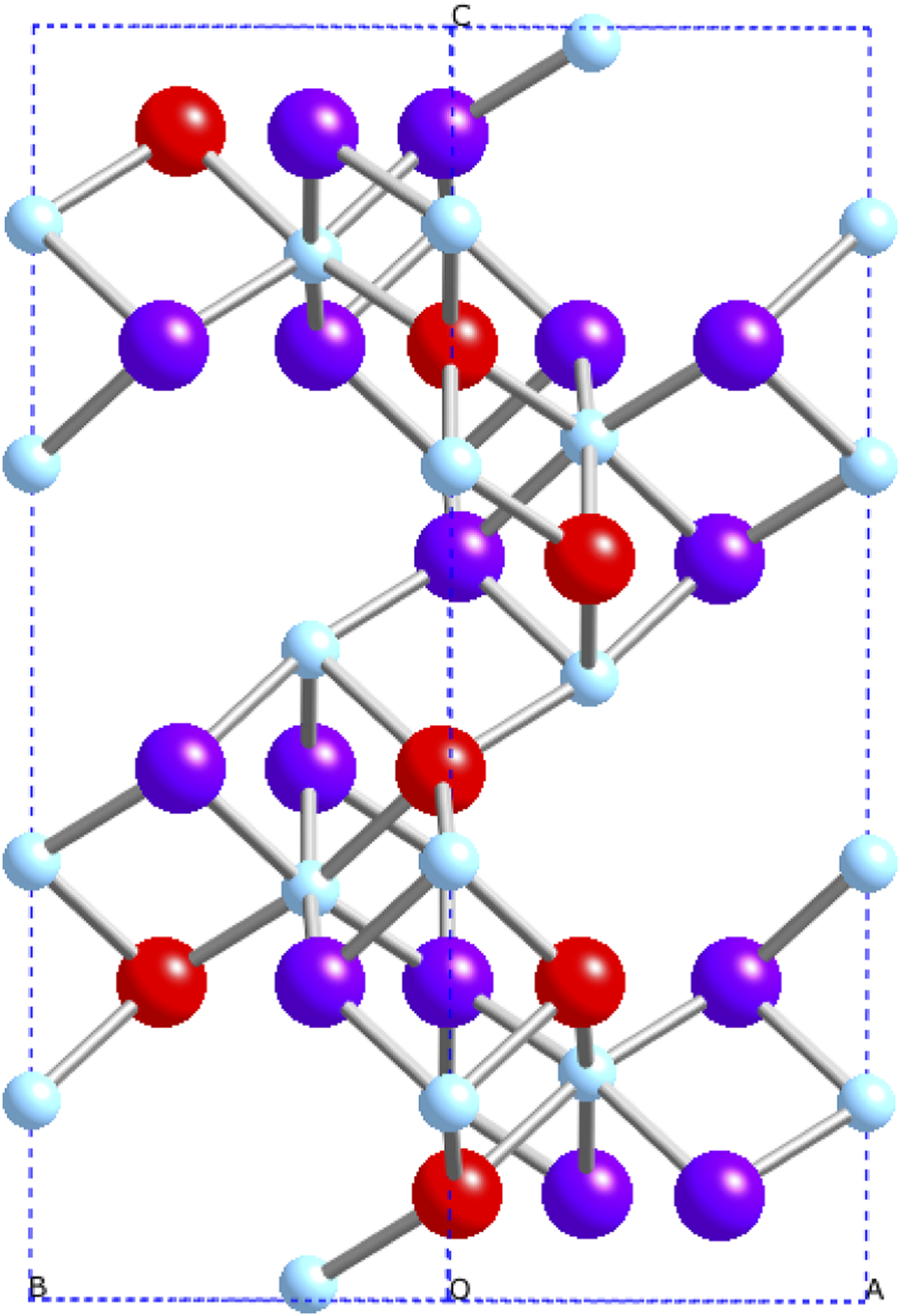} \\
(d): Ti$_2$N$_2$O-IV & (e): Ti$_2$N$_2$O-V &(f): Ti$_2$N$_2$O-VI \\
\includegraphics[width=1in]{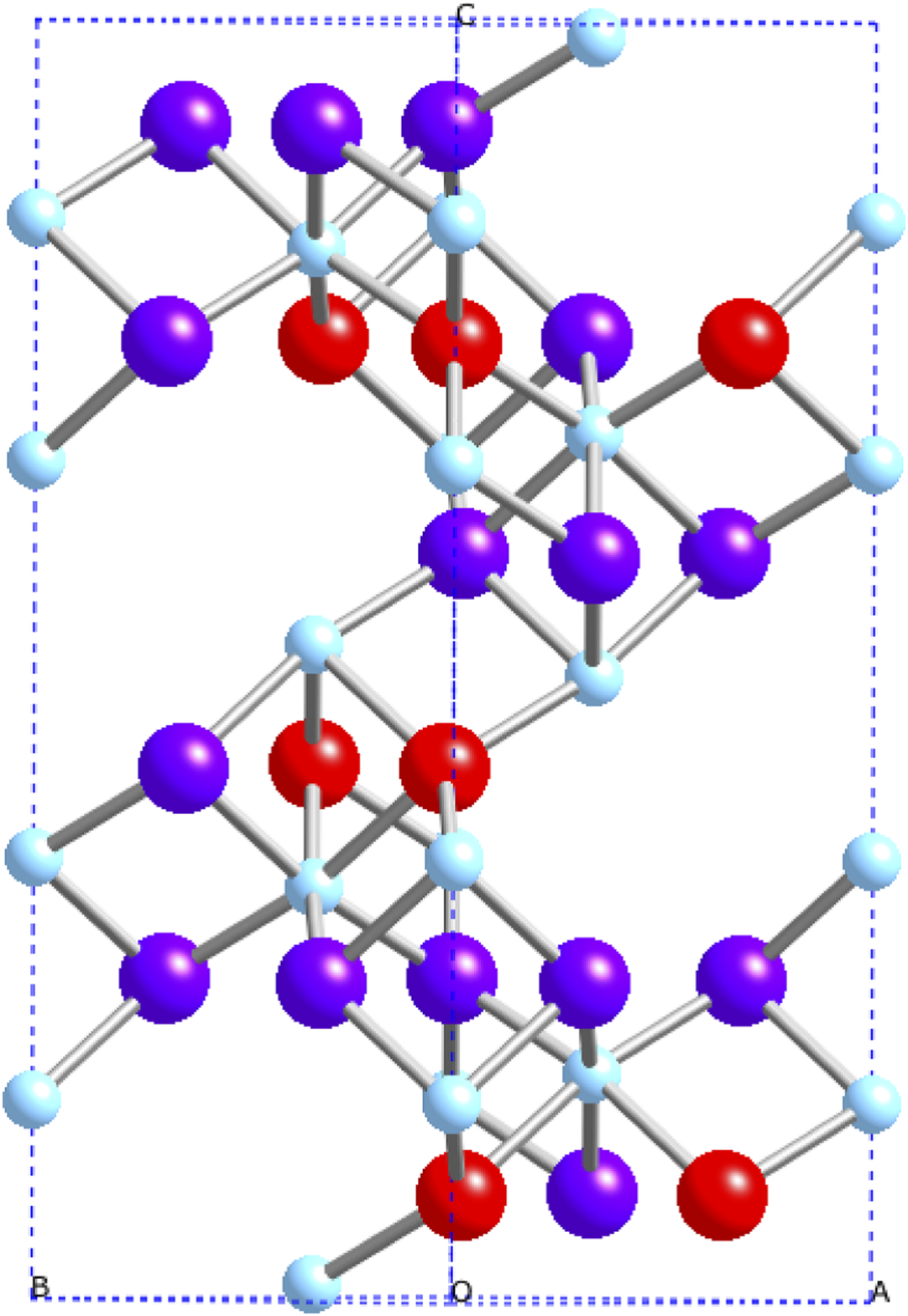} &
\includegraphics[width=1in]{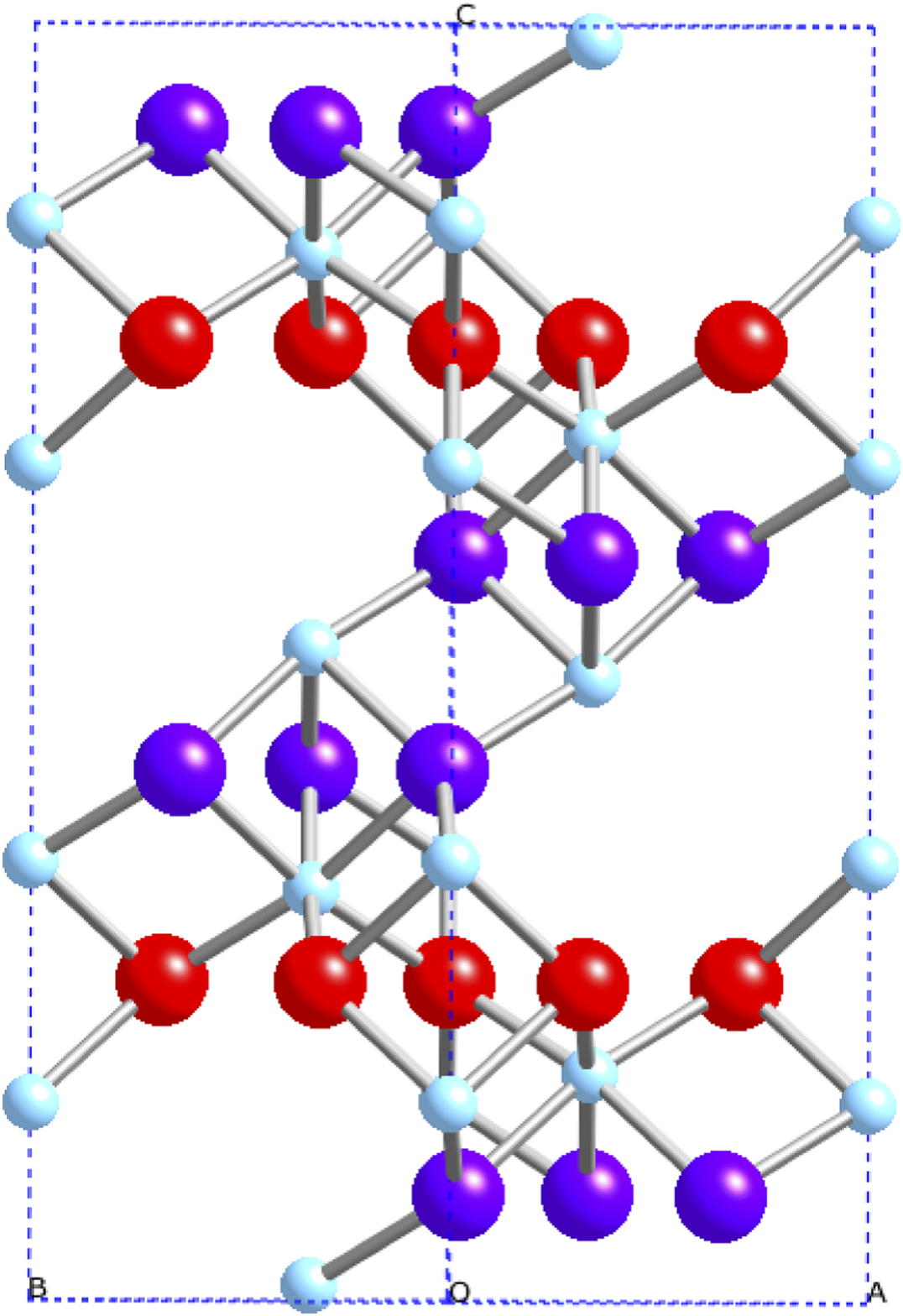} &
\includegraphics[width=1in]{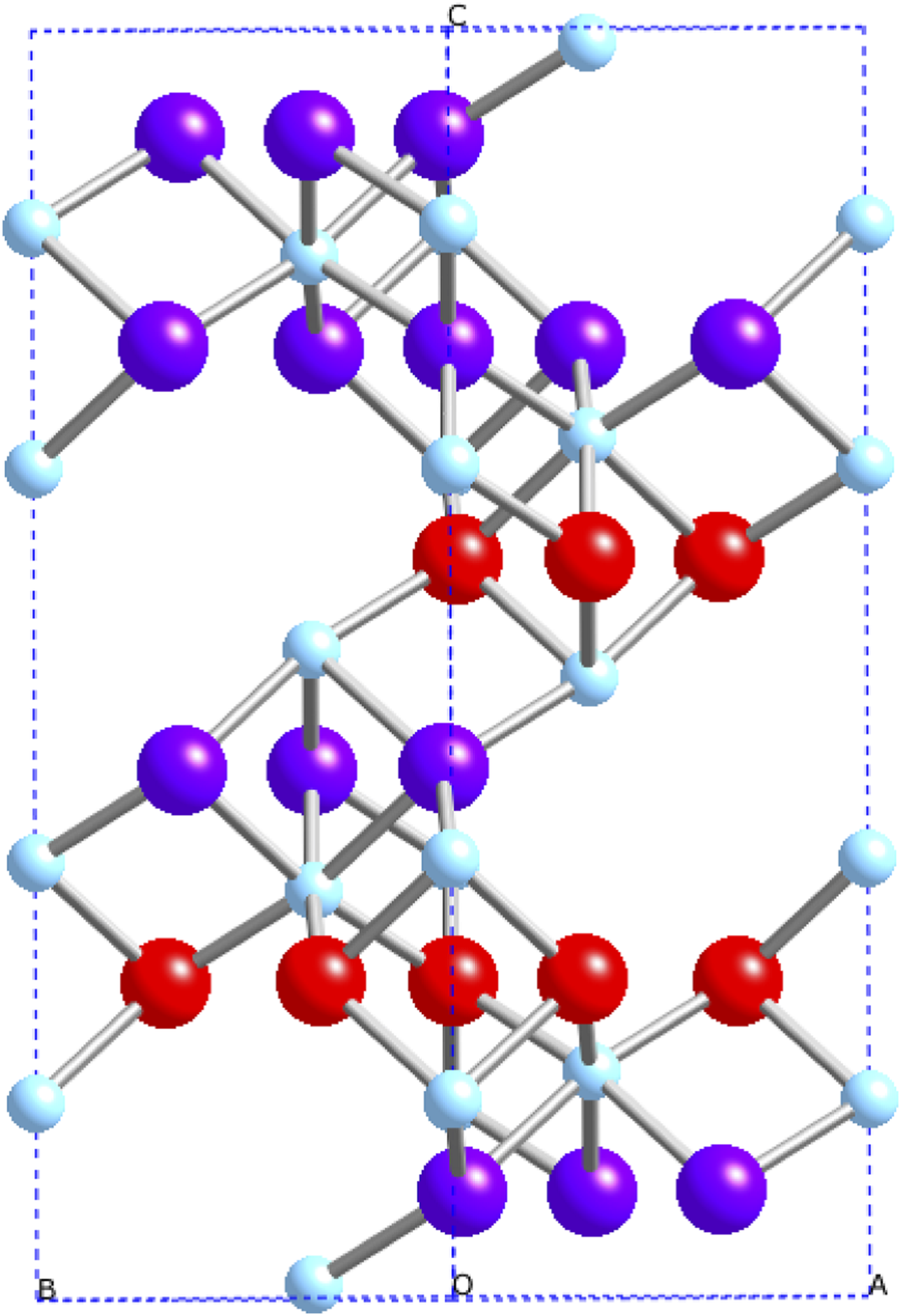} \\
(g): Ti$_2$N$_2$O-VII &&\\
\includegraphics[width=1in]{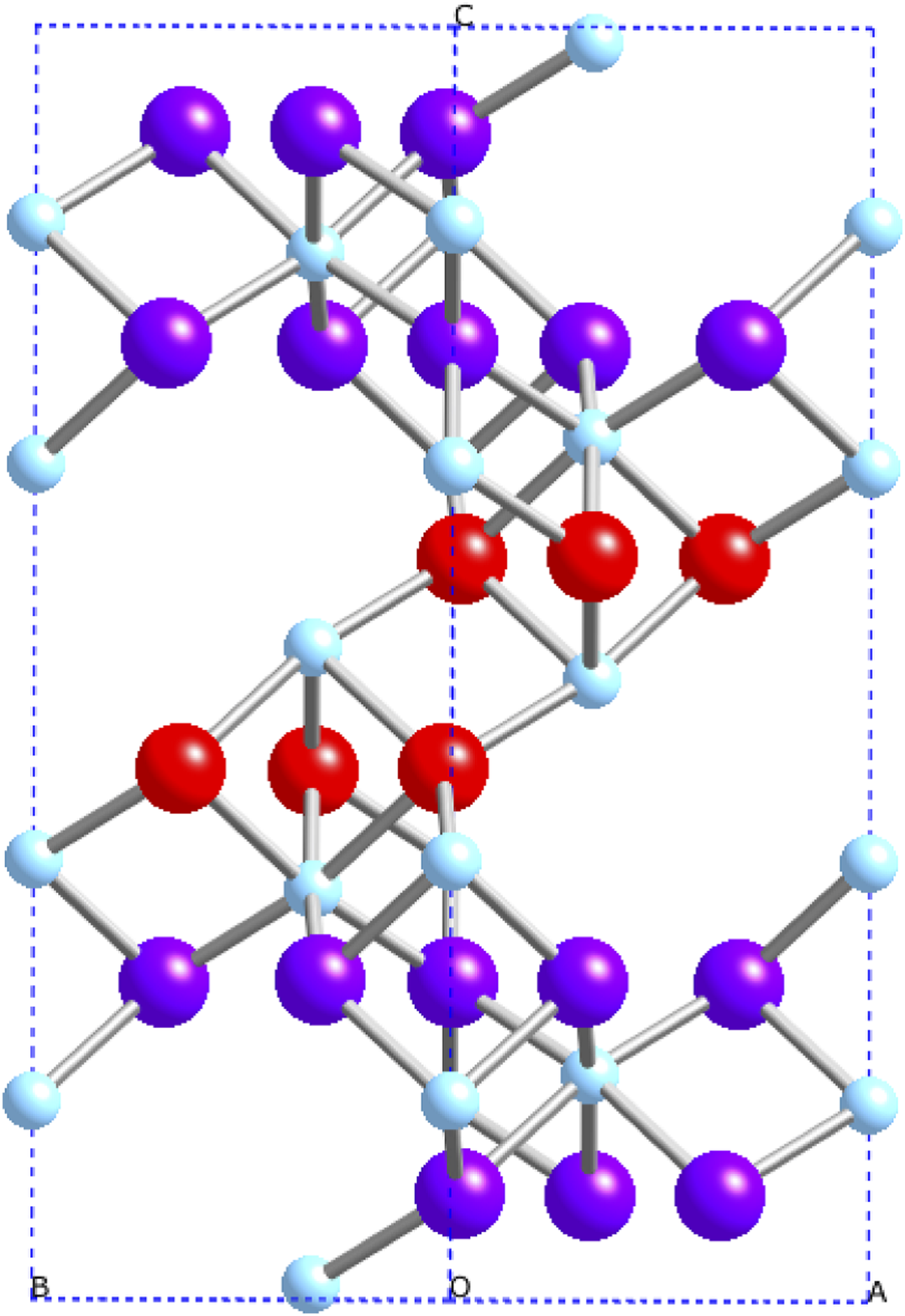} &\includegraphics[scale=0.03]{Ti-O-N_legend_2.eps}  &
\end{tabular}
\caption{(Color online) Side views of seven proposed structural models for titanium oxynitride Ti$_2$N$_2$O. These figures are projected along the $\mathbf{a}+\mathbf{b}$ direction.  }\label{Ti2N2O-st}
\end{center}
\end{figure}

We use hexagonal conventional unit cell of Ti$_2$O$_3$ with twelve titanium and eighteen
oxygen atoms
as a starting point to obtain structure of titanium oxynitride
Ti$_2$N$_2$O.  We constructed seven representative
structural models
of Ti$_2$N$_2$O by replacing twelve out of eighteen oxygen atoms with nitrogen.
Afterwards we relax the structure until it reaches total energy
minimum.
Seven structures of Ti$_2$N$_2$O we considered are shown in
Fig.~\ref{Ti2N2O-st} and we label them as Ti$_2$N$_2$O-I through
Ti$_2$N$_2$O-VII in the order of the energetic stability (Ti$_2$N$_2$O-I is the most stable).   
We find that the differences in the formation energy of these seven models are well correlated with the
 nitrogen and oxygen coordination of titanium atoms. Four out
of seven models have all titanium atoms surrounded by four nitrogen
and two oxygen atoms.  
These four models have a very similar calculated total energy (it varies at most 73~meV per
 Ti$_2$N$_2$O formula unit).
Remaining three models
are significantly less energetically stable (up to 239~meV per Ti$_2$N$_2$O formula unit)
which we relate to the fact that some of their titanium atoms are surrounded with six
nitrogen and no oxygen atoms, as shown in Table~\ref{energetic}.

\begin{table}[htbp]
\caption{
Total energy ($E_{\rm tot}$) in meV per Ti$_2$N$_2$O formula unit relative
to the most stable configuration (model I).  We also categorize the titanium atom by 
its coordination (the numbers of neighboring oxygen and nitrogen atoms) and list the number of each 
category of titanium atoms in the computational cell. There are twelve titanium atoms per cell
in total. 
The most stable structures (with lowest $E_{\rm tot}$) have all titanium
atoms surrounded with four nitrogen and two oxygen atoms.  
On the other hand, titanium atoms surrounded with six nitrogen 
and no oxygen atoms are energetically unfavored.}
\label{energetic}
\begin{ruledtabular}
\begin{tabular}{cccm{0.5cm}m{0.5cm}m{0.5cm}m{0.5cm}}
Model & $E_{\rm tot}$ & \multicolumn{5}{c}{Titanium coordination }\\ 
\cline{3-7}
      & (meV/Ti$_2$N$_2$O) & \#N & 6 &  4 & 3 &  0 \\ 
      &       & \#O   & 0 &  2 & 3 &  6 \\ 
\hline
I     &   0   & &  0 &   12 &   0 &    0 \\
II     &  21   & &  0 &   12 &   0 &    0 \\
III    &  21   & &  0 &   12 &   0 &    0 \\
IV   &  73   & &  0 &   12 &   0 &    0 \\
V   & 102   & &  4 &    0 &   8 &    0 \\
VI    &  130   & &  4 &    0 &   8 &    0 \\
VII   & 239   & &  6 &    0 &   4 &    2 \\
\end{tabular}
\end{ruledtabular}
\end{table}

\section{Nitrogen substitution energy}

Now we compare energy required to substitute a nitrogen
atom into TiO$_2$ and Ti$_2$O$_3$.  Since there are no oxygen
vancancies in TiO$_2$ we expect that energy required for substitution
in TiO$_2$ will be significantly larger than that in Ti$_2$O$_3$.

We define the substitutional energy $E_{\rm s}$ per single nitrogen atom in the case of TiO$_2$ and Ti$_2$O$_3$ as,
\begin{widetext}
\begin{align}
E_{\rm s} ({\rm Ti O}_2) &= E_{\rm tot}({\rm Ti}_{16}{\rm O}_{31}{\rm N}) - 16E_{\rm tot}({\rm TiO_2}) + \frac{1}{2}E_{\rm tot}({\rm O_2}) - \frac{1}{2}E_{\rm tot}({\rm N_2})\label{formNdoped} \\
E_{\rm s} ({\rm Ti}_2{\rm O}_3) &= \frac{1}{2} \left[E_{\rm tot}({\rm Ti_2N_2O}) - E_{\rm tot}({\rm Ti_2O_3}) + E_{\rm tot}({\rm O_2}) - E_{\rm tot}({\rm N_2})\right]. \label{formTi2N2O}
\end{align}
\end{widetext}
Where $E_{\rm tot}({\rm TiO_2})$ and $E_{\rm tot}({\rm Ti_2O_3})$ are calculated total energies of pristine phases while $E_{\rm tot}({\rm O_2})$ and $E_{\rm tot}({\rm N_2})$ are calculated total energies of molecular oxygen and nitrogen.

Calculated values of substitutional energy $E_{\rm s}$ are shown in
Table~\ref{energeticboth} for the cases of TiO$_2$ and Ti$_2$O$_3$. In the case of TiO$_2$, $E_{\rm s}$ values are 5.74~eV for the rutile phase and 5.89~eV for the anatase phase. In the case of Ti$_2$O$_3$, on the other hand, $E_{\rm s}$ values are significantly lower, in the range of 3.30--3.42~eV. Therefore, nitrogen substitution into Ti$_2$O$_3$ is
energetically more favorable than nitrogen doping into TiO$_2$, 
as is expected from the titanium valency of +IV in Ti$_2$N$_2$O. 
This result shows that Ti$_2$N$_2$O could be created in the laboratory and 
the existence of Ti$_2$N$_2$O molecules\cite{Ti2N2Omol} also supports this conclusion.

\begin{table}[htbp]
\caption{
Summary of results for a conventional substitution of nitrogen
into TiO$_2$ as well as substitution of nitrogen
into Ti$_2$O$_3$ forming oxynitride Ti$_2$N$_2$O.  We show substitutional energy per single nitrogen atom
$E_{\rm s}$, electron gap, conduction band minimum (CBM) and valence band
maximum (VBM).
CBM and VBM are aligned in energy by using the oxygen
2$s$ core level in each system as a reference. After aligning the 
energy levels, we introduce the scissor-shift by adding $\Delta E_{\rm g}/2$ to 
CBM and subtracting it from VBM. 
$\Delta E_{\rm g}$ is obtained from the GW calculation in 
pristine cases and is approximated by Eq.~(\ref{linearmodel}) in nitrogen-substituted cases. 
Values without scissor-shifts are listed in the parentheses as a reference.
For both values with and without scissor-shift, we take VBM of pristine rutile as zero in energy.
Parent Ti$_2$O$_3$ phase is metallic within DFT so we only list in the Table location of the Fermi level (within DFT and without scissor-shift).}
\label{energeticboth}
\begin{ruledtabular}
\begin{tabular}{lcccc}
  & $E_{\rm s}$ & Gap & CBM & VBM \\ & (eV) & (eV) & (eV) & (eV) \\
\hline
\multicolumn{5}{l}{Pristine TiO$_2$} \\
\quad \quad Rutile   &      &  3.14   &  3.14  &   0.00   \\
                     &      & (1.87)  & (1.87) &  (0.00)  \\
\quad \quad Anatase  &      &  3.55   &  3.11  & $-0.45$   \\
                     &      & (2.13)  & (1.76) &($-0.37$)  \\
\multicolumn{5}{l}{TiO$_2$ with 1/32 N substitution} \\
\quad \quad Rutile   & 5.74 &  3.11  &  3.17  &   0.06  \\
                     &      & (1.86) & (1.91) &  (0.05)\\
\quad \quad Anatase  & 5.89 &  2.98  &  3.00  &   0.03  \\
                     &      & (1.78) & (1.77) & ($-0.01$)\\
\multicolumn{5}{l}{Pristine Ti$_2$O$_3$} \\
 &      & \multicolumn{3}{c}{{\rm \it Fermi level at 3.27 eV}}  \\
\multicolumn{5}{l}{Ti$_2$N$_2$O oxynitride} \\

\quad \quad I       & 3.30 &1.81   &3.15  &1.34  \\
                    &      & (1.08)& (2.15) &  (1.07)\\

\quad \quad II      & 3.31 &2.19   &3.16   &0.96  \\
                    &      & (1.31)& (2.08) & (0.77)\\

\quad \quad III     & 3.31 &2.11   &3.12   &1.01   \\
                    &      & (1.26)& (2.06) &  (0.80)\\

\quad \quad IV      & 3.34 &2.06   &3.28   &1.22  \\
                    &      & (1.23)&  (2.23) &  (1.00)\\

\quad \quad V       & 3.35 &2.32   &3.35   &1.03  \\
                    &      & (1.39)&  (2.25) &  (0.86)\\

\quad \quad VI      & 3.37 &2.01    &3.26   &1.25  \\
                    &      & (1.20) & (2.22) & (1.02)\\

\quad \quad VII     & 3.42 &1.99    &3.74    &1.76  \\
                    &      &  (1.19)&  (2.71) &  (1.52)\\

\end{tabular}
\end{ruledtabular}
\end{table}

\section{Electronic structure}

We now
discuss its effect on the electronic structure.  As is well known,
Kohn-Sham energies in the DFT calculation tend to underestimate
electronic gaps in most semiconductors. \cite{hybertsen-louie-1,hybertsen-louie-2,gap1,gap2}  Therefore, we use
here GW approximation to obtain improved electron gaps.
Since these calculations are rather computationally demanding, we
performed them on pristine TiO$_2$ rutile and anatase and use them to infer GW corrections in nitrogen-substituted cases.

\subsection{Pristine TiO$_2$ and Ti$_2$O$_3$} \label{GWcal}

\begin{figure*}[htbp]
\begin{center}
\begin{tabular}{cc}
(a): Rutile& (b): Anatase\\
\includegraphics[scale=0.6]{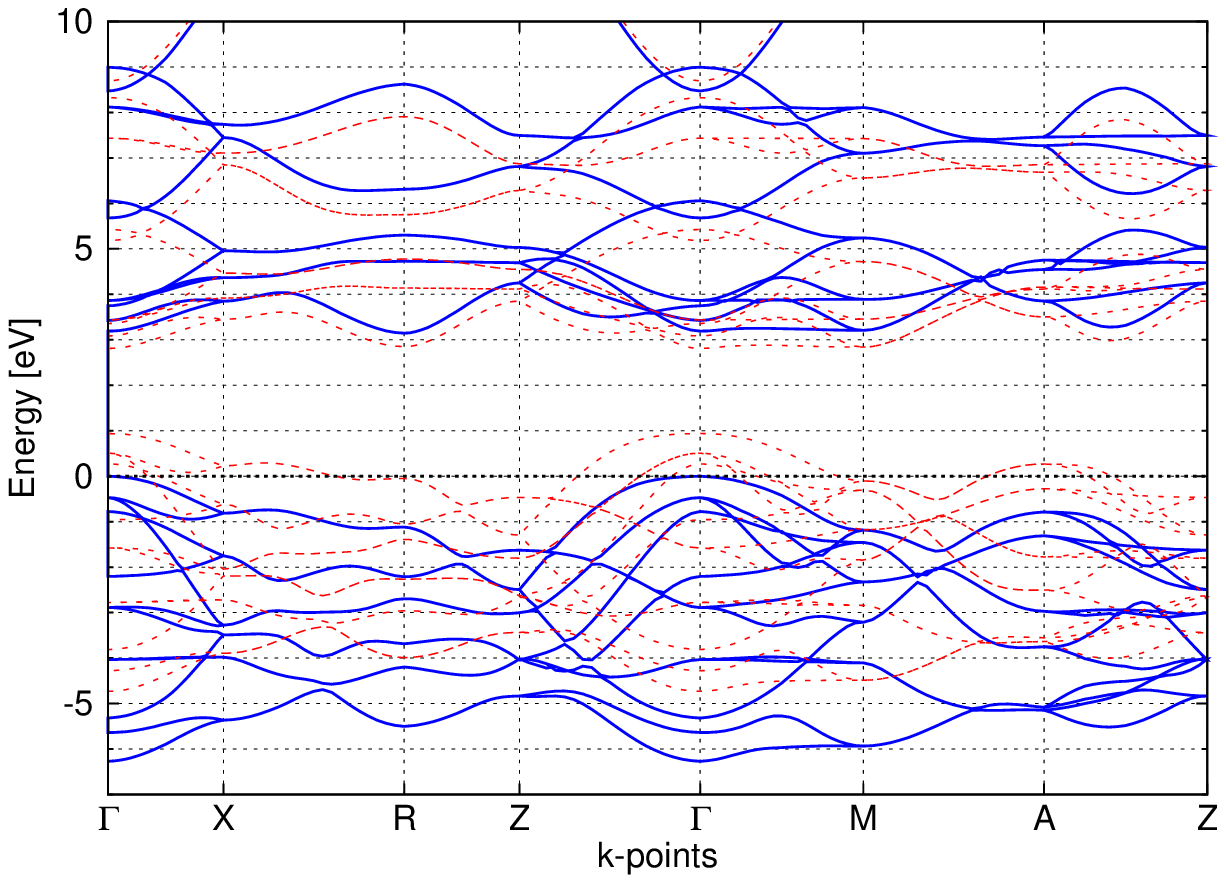} &
\includegraphics[scale=0.6]{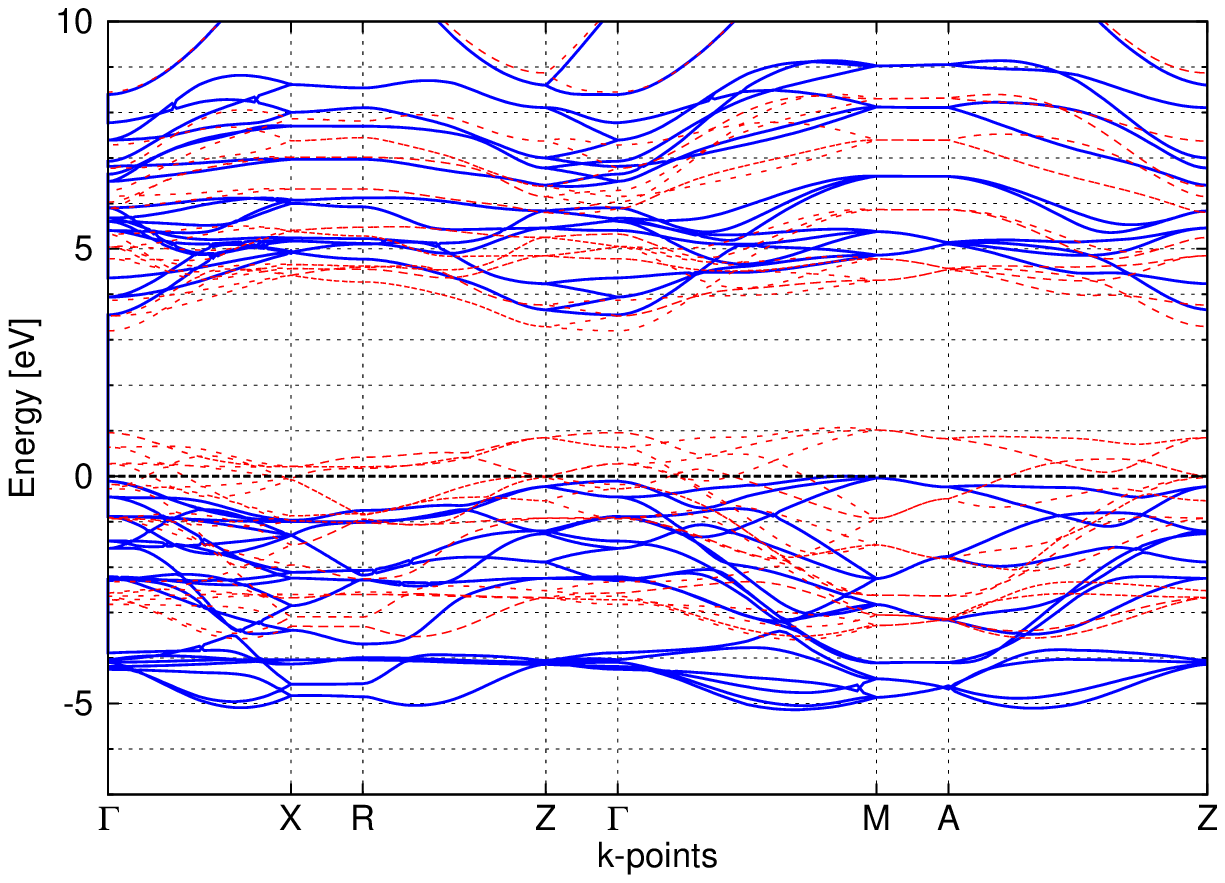}
\end{tabular}
\caption{(Color online) DFT (dashed red line) and GW-corrected (solid blue line) band
  structures of pristine rutile and anatase. The valence band maximum calculated with GW correction is
  taken as zero.}\label{rutile-anatase}
\end{center}
\end{figure*}

Figure~\ref{rutile-anatase} shows DFT (dashed red line)
and GW-corrected (solid blue line) band structures of pristine rutile
and anatase TiO$_2$. 
We find that the GW-corrected band structure is nearly rigidly opened-up
compared to the DFT band structure (scissor-shifted).

We find that the rutile TiO$_2$ has a direct band gap of 1.87~eV at the
$\Gamma$ point within a DFT calculation.  Inclusion of the GW
correction changes this direct band gap to 3.19~eV
and produces a somewhat smaller indirect $\Gamma$-to-R gap of
3.14~eV. This result shows that the scissor-shift due to the GW
correction is not completely rigid.
 Similar result was obtained in a recent GW
study as well.\cite{Andrei}  Nevertheless, this variation is relatively small.

In the case of anatase TiO$_2$ we find conduction-band minimum (CBM)
at the $\Gamma$ point but the valence-band maximum (VBM) on the
$\Gamma$--M line, close to the M point, both in DFT and GW-corrected results.
Therefore, the band gap in anatase is indirect. Its magnitude is 2.13~eV within DFT and 3.55~eV
with the GW correction.

In the case of pristine Ti$_2$O$_3$, we find it to be metallic within
the DFT approximation.  
Therefore, 
further detailed studies are needed both experimentally and theoretically 
to discuss the small charge-ordering gap found in previous experiments.\cite{ti2o3-1,ti2o3-2,ti2o3-3,ti2o3-4,MITreview,ti2o3-9}
On the other hand, 
oxynitride Ti$_2$N$_2$O 
does not possess such difficulty and
is insulating with the substantial gap even within DFT
as expected from the electron counting argument, which is shown in the following subsection.

\begin{figure}[htbp]
\begin{center}
\includegraphics[scale=0.6]{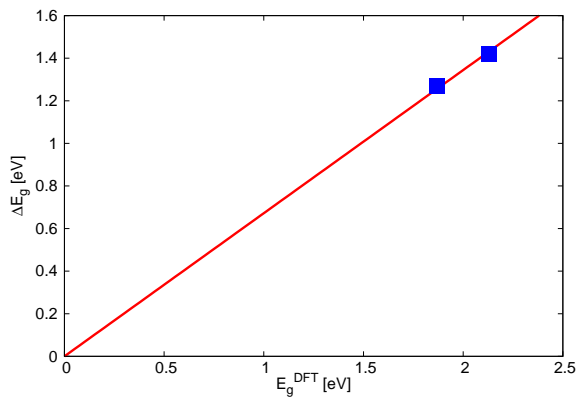} 
\caption{(Color online) The GW corrections to the band gap, $\Delta E_{\rm g}$, are plotted with respect to the DFT band gap $E_{\rm g}^{\rm DFT}$ both for pristine rutile and anatase.}\label{DeltaEg}
\end{center}
\end{figure}

Here, we analyze GW corrections to the band gap in pristine rutile and anatase TiO$_2$. 
We label GW-corrected and DFT band gaps as $E_{\rm g}^{\rm GW}$ and $E_{\rm g}^{\rm DFT}$, 
respectively, and define the GW correction to the band gap as $\Delta E_{\rm g} \equiv E_{\rm g}^{\rm GW} - E_{\rm g}^{\rm DFT}$.
We plot the $\Delta E_{\rm g}$ with respect to $E_{\rm g}^{\rm DFT}$ both for rutile and anatase TiO$_2$ in Fig.~\ref{DeltaEg}. 
As one can see from the figure, two data point for rutile and anatase are almost collinear with the original point, 
indicating that $\Delta E_{\rm g}$ is approximately proportional to $E_{\rm g}^{\rm DFT}$ 
in titanium oxide systems. The proportional trend of the GW correction to the band gap has also been found in the 
previous literatures. \cite{godby-gw, louie-gw, kotani-gw}
Once we assume the proportionality between $\Delta E_{\rm g}$ and $E_{\rm g}^{\rm DFT}$, 
we obtain
\begin{equation}
\Delta E_{\rm g} \approx 0.67 E_{\rm g}^{\rm DFT} \label{linearmodel}
\end{equation}
from the least squares fitting. 
We use Equation~(\ref{linearmodel}) to approximate the scissor-shifts in nitrogen-substituted cases.

\subsection{Nitrogen-substituted cases}

Now we are ready to discuss the electronic structure of nitrogen-substituted 
systems.  
We list the electron gap, the conduction band maximum (CBM), and the valence band 
maximum (VBM) calculated based on DFT in Table~\ref{energeticboth} 
for the cases of TiO$_2$-derived and Ti$_2$O$_3$-derived systems. 
In order to align the energy levels, we use the oxygen $2s$ core level in each system as the 
reference energy instead of the vacuum level since the calculation of the vacuum level 
requires detailed analysis of surface conditions. 
Although energy-level alignment referenced to the oxygen $2s$ level is an approximation, 
we expect that it works well in this case since it reproduces the rutile/anatase band alignment
in pristine TiO$_2$ both measured by XPS\cite{alignmentexp} and predicted by DFT\cite{alignmentexp, alignmentcomp} as well as the band-edge shift 
induced by nitrogen doping into rutile or anatase predicted in a previous study.\cite{valentin}

Our results also show that CBM and VBM of anatase are lower than those of rutile by 0.03~eV and 0.45~eV, respectively.

After energy level alignment we introduce the GW scissor-shift correction by adding $\Delta E_{\rm g}/2$ to CBM and subtracting it from VBM.  Therefore total gap is increased by $\Delta E_{\rm g}$.

Aligned and scissor-shifted energy levels are reported in Table~\ref{energeticboth}. We also list the values without the scissor-shifting in the parentheses for references. 
In the case of nitrogen-doped TiO$_2$, we define the gap as the energy difference between
CBM and the Fermi energy (Fermi energy is treated as VBM) since the Fermi energy is on the 
localized impurity state as discussed later.

Figure~\ref{fig:pdos} compares projected
density of states into titanium $d$, oxygen $p$, and nitrogen $p$ orbitals together with the total density of states
calculated based on DFT. 
In this figure, energy levels are aligned and both conduction and valence bands are GW scissor-shifted in the same way as in Table~\ref{energeticboth}.

We now discuss  these results in more detail.

\subsubsection{Nitrogen substituted TiO$_2$}

\begin{figure*}[htbp]
\begin{center}
\begin{tabular}{cccccc}
&&&&&\includegraphics[width=0.5in]{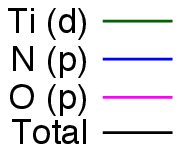}\\ (a):
Rutile & (b): Anatase &(c): N-doped rutile &(d): N-doped anatase &
(e): Ti$_2$O$_3$ &(f)Ti$_2$N$_2$O-I \\
\includegraphics[width=1in]{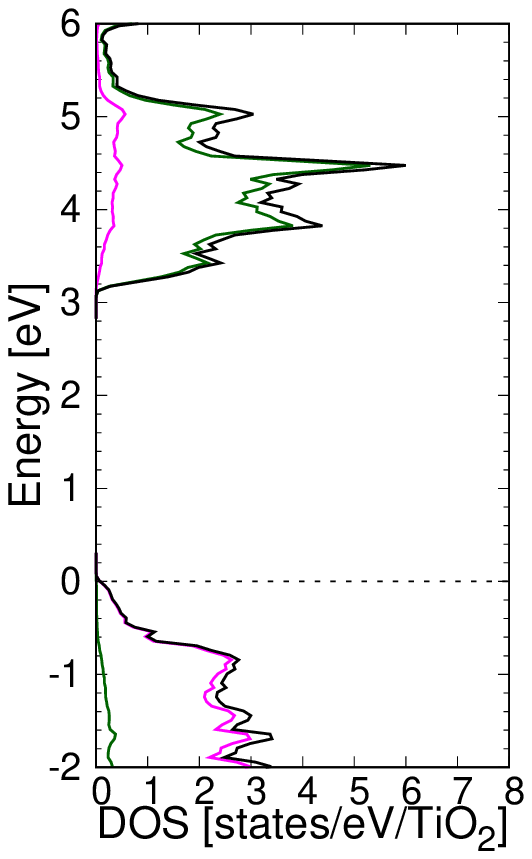} &
\includegraphics[width=1in]{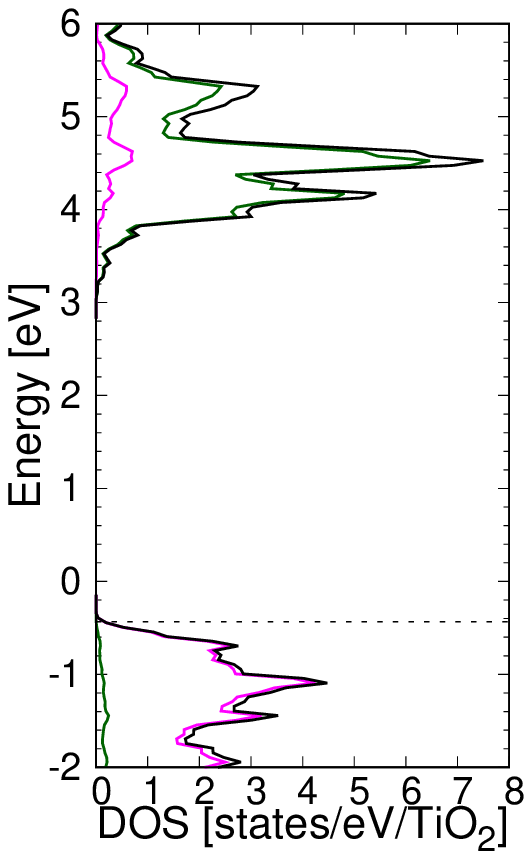} &
\includegraphics[width=1in]{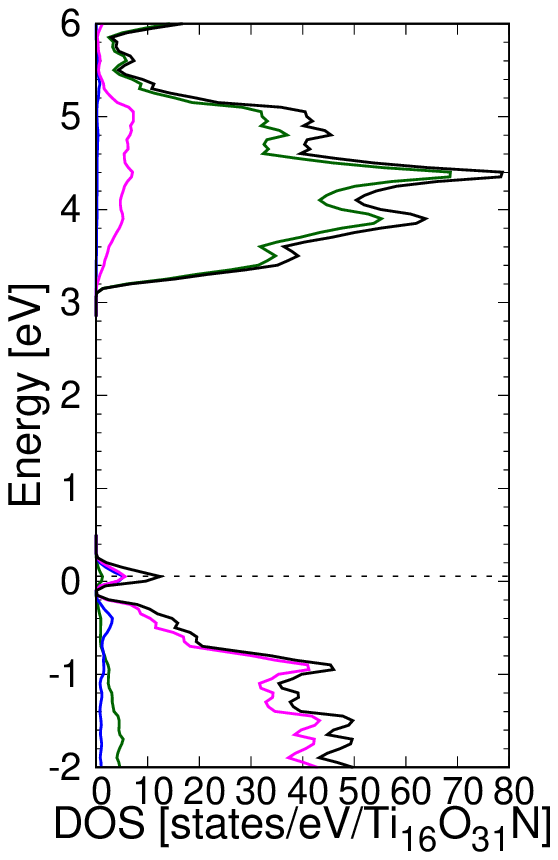} &
\includegraphics[width=1in]{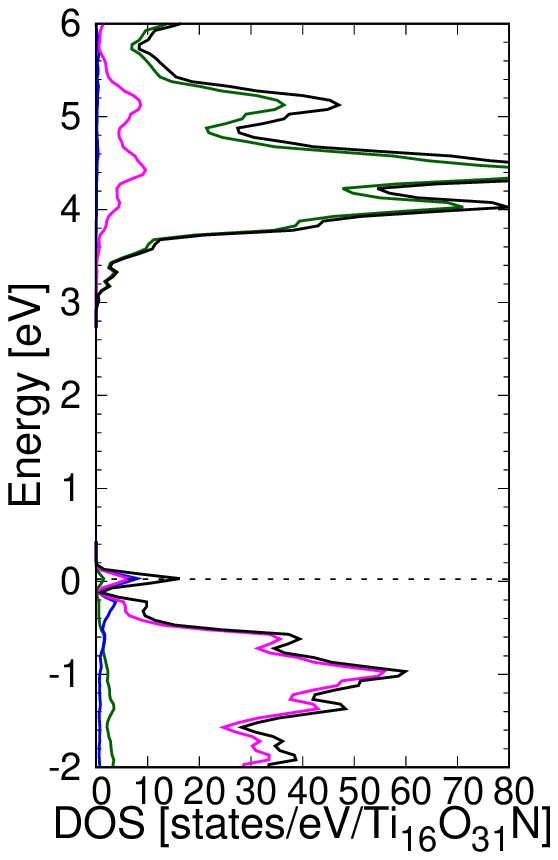} &
\includegraphics[width=1in]{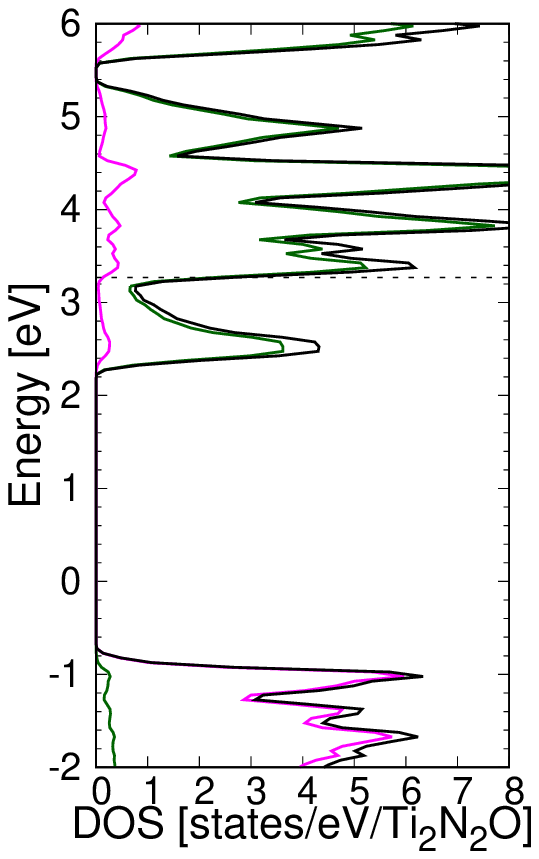} &
\includegraphics[width=1in]{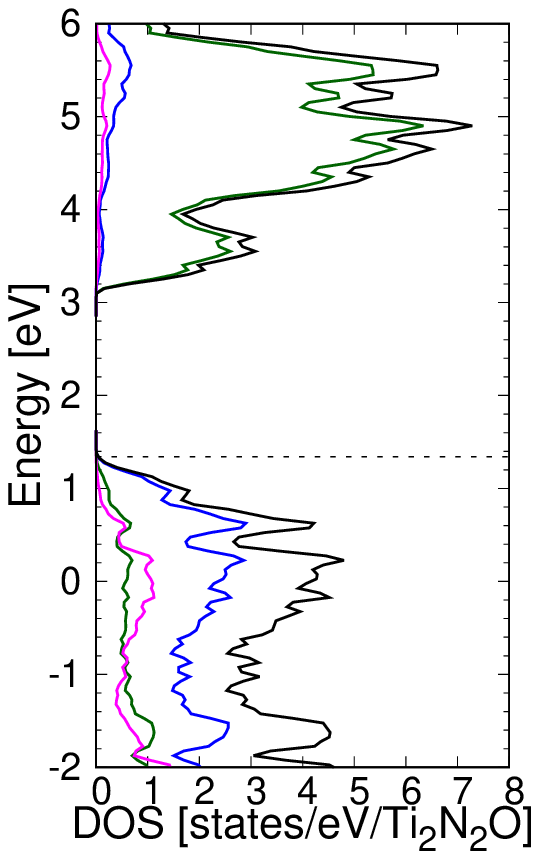} \\
(g): Ti$_2$N$_2$O-II & (h): Ti$_2$N$_2$O-III &(i): Ti$_2$N$_2$O-IV &
(j): Ti$_2$N$_2$O-V & (k): Ti$_2$N$_2$O-VI &(l): Ti$_2$N$_2$O-VII \\
\includegraphics[width=1in]{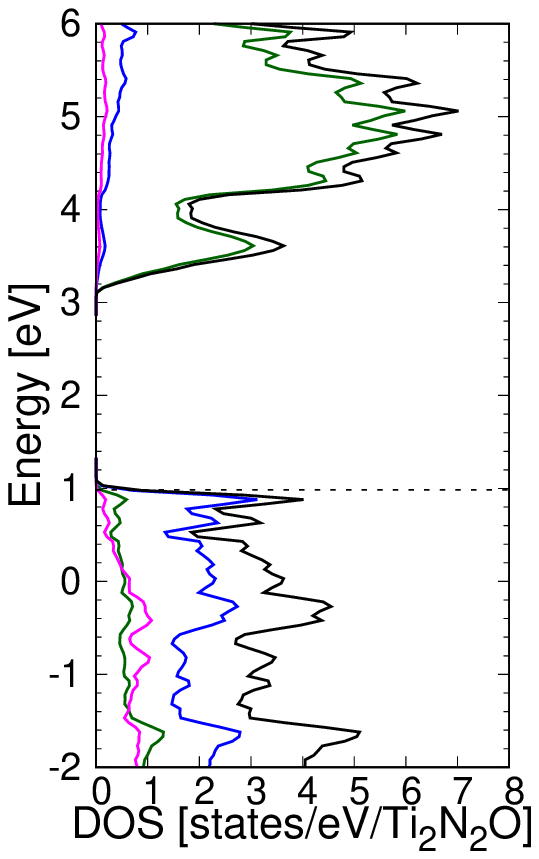} &
\includegraphics[width=1in]{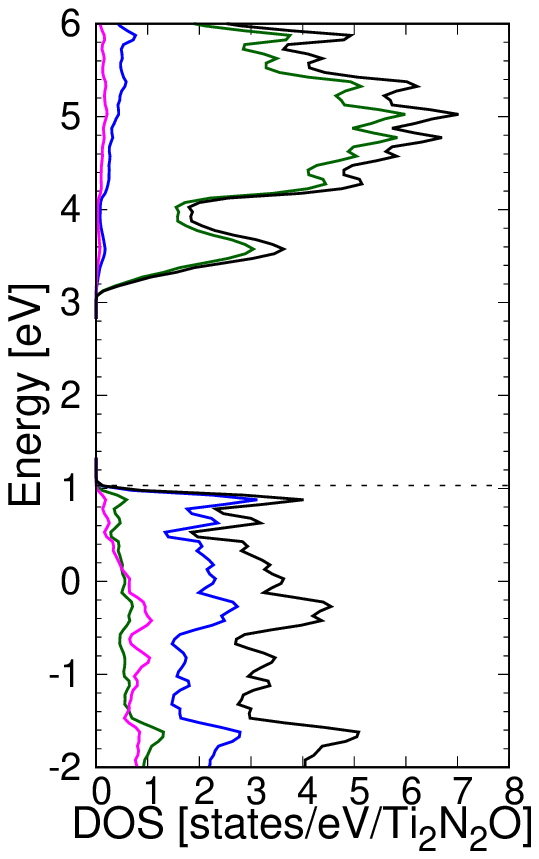} &
\includegraphics[width=1in]{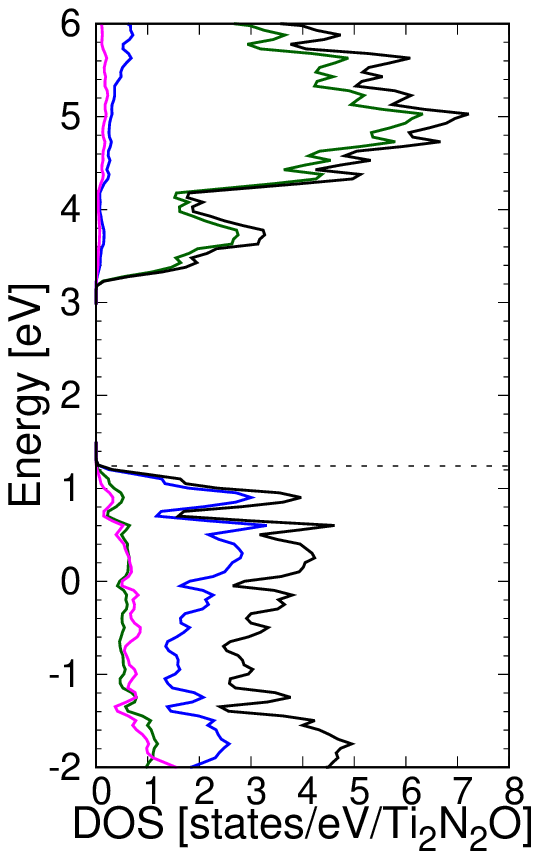} &
\includegraphics[width=1in]{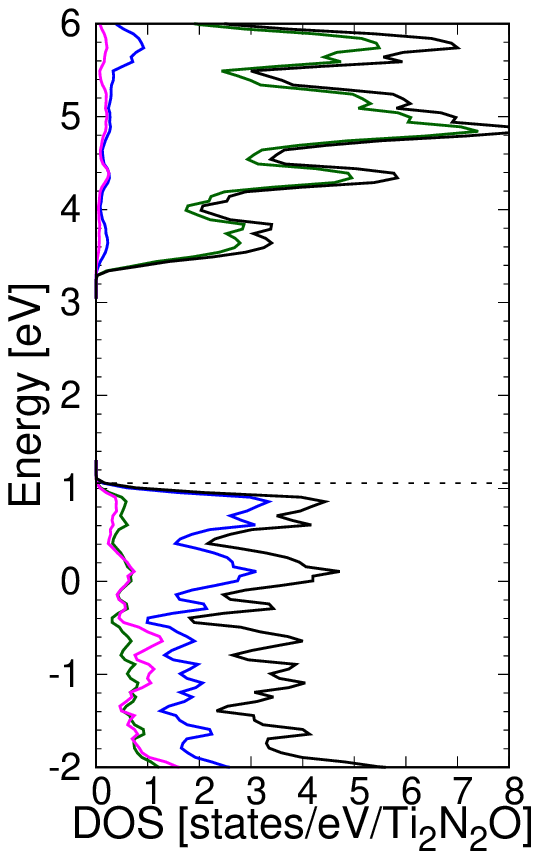} &
\includegraphics[width=1in]{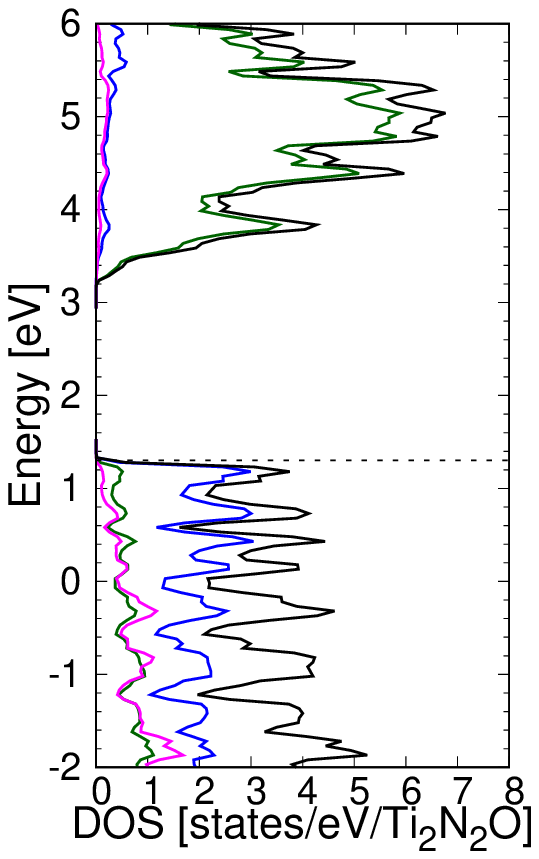} &
\includegraphics[width=1in]{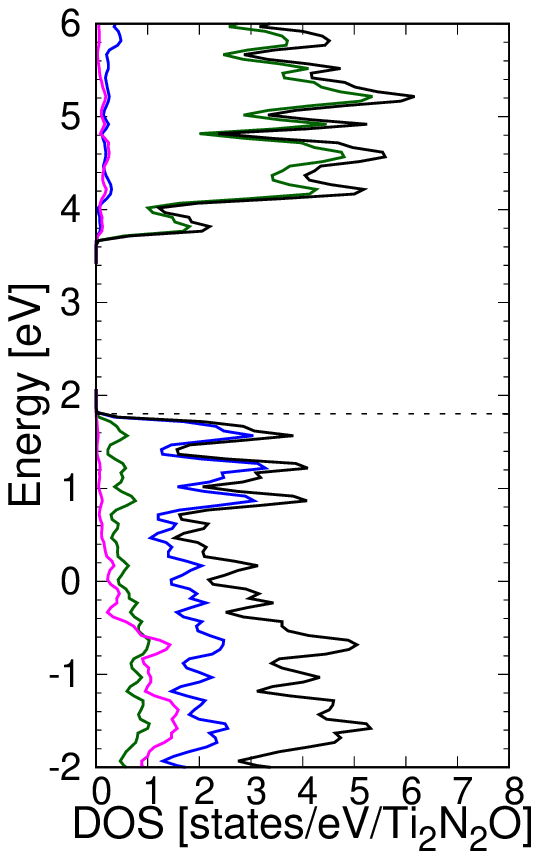} \\
\end{tabular}
\caption{(Color online) Projected density of states of all  systems that we studied. Dotted 
straight lines indicate VBM (or the Fermi energy for
metals). Energy levels are aligned and 
scissor-shifted  as in Table \ref{energeticboth}.}
\label{fig:pdos}
\end{center}
\end{figure*}

Substituting nitrogen into either rutile or anatase TiO$_2$ induces localized
states just above the valence band maximum.  These localized states are 
0.25~eV above valence bands in rutile and 0.20~eV in anatase.
In both cases Fermi level is crossing the localized band.

However, nitrogen substitution not only introduces localized states, but also shifts valence and
conduction bands relative to the undoped case.  As a result, upon nitrogen substitution,
the gap of rutile TiO$_2$ shifts from 3.14~eV to 3.11~eV while the gap of anatase TiO$_2$ changes
from 3.55~eV to 2.98~eV.
Similar band-edge shifts are also predicted in a previous DFT study. \cite{valentin}

As can be seen from Fig.~\ref{fig:pdos} in pristine and nitrogen substituted rutile and 
anatase TiO$_2$ the valence band is composed mostly of oxygen $p$ states, while the isolated impurity
band is dominated by the nitrogen $p$ states.

\subsubsection{Nitrogen substituted Ti$_2$O$_3$: Ti$_2$N$_2$O}

Now we discuss electronic structure of seven nitrogen-substituted
(Ti$_2$N$_2$O) models.  We find that all seven models are semiconducting as expected from the 
electron-counting rule. 

The calculated band-gap values of Ti$_2$N$_2$O models range from 1.81 to 2.32~eV.
Among the seven Ti$_2$N$_2$O models, Ti$_2$N$_2$O-I has the
smallest band gap.  This model is also energetically the most stable one.
The band gap of Ti$_2$N$_2$O-I is therefore significantly  smaller than that of  
pristine rutile, anatase, or even nitrogen-doped rutile and anatase.  In addition, in 
Ti$_2$N$_2$O, the top of the conduction band is not formed by the localized band as in the case
of nitrogen substituted rutile and anatase.

However, as discussed earlier, photocatalytic activity of TiO$_2$ depends not only on the
electronic band gap but it depends also on energy level alignment.

We find that the top of the valence band
of the seven Ti$_2$N$_2$O models are higher than that of pristine rutile TiO$_2$
by 0.96--1.76 eV.  This alignment is therefore more preferable for water splitting purposes.

Conduction band minimum of most Ti$_2$N$_2$O models are higher than that of 
pristine rutile TiO$_2$.  The most stable model (I) had nearly the same conduction band maximum
as pristine rutile TiO$_2$.

Therefore, the reduction of band-gap values in Ti$_2$N$_2$O is achieved mostly by the 
upward shift of VBM, not by the downward shift of CBM. 
Therefore Ti$_2$N$_2$O possesses suitable band-edge position for photocatalytic 
water decomposition. 
In all the Ti$_2$N$_2$O models, the upper part of the valence band is composed of mainly 
nitrogen 2$p$ states, as shown in Fig.~\ref{fig:pdos}. This result shows the effectiveness of nitrogen substitution to 
shift up the VBM and thus to form a narrower band gap. The amount of band-edge shift in Ti$_2$N$_2$O-I, 
the most stable model, is highly notable because it is the most likely candidate among the seven models.
Its VBM and CBM shifts from those of pristine rutile TiO$_2$ are 1.34~eV and 0.01~eV in the upward direction, respectively.

\section{Summary}

In this paper, we have presented a general approach to obtain a variety of purely insulating
titanium oxynitride compounds Ti$_n$N$_2$O$_{2n-3}$ and we conducted
detailed analyses in the cases with the highest nitrogen concentration ($n=2$). 
The resulting oxynitride, Ti$_2$N$_2$O, is predicted to have
suitable band-edge position for photocatalytic water decomposition.  Its band gap (1.81~eV) 
is much smaller in pristine TiO$_2$ (3.14--3.55~eV).  Importantly, most of the band gap
reduction originates from the upward shift in energy of the valence band maximum.
Furthermore, the energy required to substitute a nitrogen atom into Ti$_2$O$_3$ to form 
Ti$_2$N$_2$O is smaller than that into TiO$_2$. 
These results show that Ti$_2$N$_2$O is a more promising material 
than nitrogen-doped TiO$_2$.
We hope that this approach would be extended to cases with lower nitrogen concentration
and to other transition-metal oxynitrides.

\begin{acknowledgments}
This work was supported from the MEXT Japan Elements Strategy Initiative to Form Core Research Center, 
JSPS KAKENHI Grant No. JP25107005, and JSPS Grant No. JP14J11856.
We also acknowledge the supports by National Science Foundation Grant No. DMR-1508412 
(which provided the DFT calculations) and the Theory of Materials Program at the Lawrence Berkeley National Lab
funded by the Director, Office of Science, Office of Basic
Energy Sciences, Materials Sciences and Engineering Division, 
U.S. Department of Energy under Contract No. DE-AC02-05CH11231 (which provided the GW calculations). 
Computational resources have been provided by the DOE at Lawrence Berkeley National Laboratory's NERSC facility.
\end{acknowledgments}

\end{document}